\documentclass[longauth,useAMS,usenatbib]{aa}

\usepackage{graphicx}
\usepackage{amsmath}   %
\usepackage{amssymb}
\usepackage{txfonts}
\usepackage{url}
\usepackage{xcolor}
\usepackage{booktabs}
\usepackage{longtable}  %
\usepackage{multirow}  %
\usepackage[utf8]{inputenc}

\newcommand{\SEM}[1]{#1}

\providecommand{\apjl}{ApJL}\providecommand{\apjs}{ApJS}
\providecommand{\aap}{A\&A}

\newcommand{\grs}{GRS\,1915$+$105}
\newcommand{\lr}{$L_{\rm R}$}
\newcommand{\lx}{$L_{\rm X}$}
\newcommand{\ergs}{erg\,s$^{-1}$}
\newcommand{\ujy}{$\mu$Jy}
\newcommand{\ujb}{$\mu$Jy\,beam$^{-1}$}

\begin{document}

\title{The retirement party of GRS\,1915$+$105: the radio fading to quiescence of the archetypal microquasar}

\titlerunning{The retirement party of \grs}

\author{S.~E. Motta\inst{1,2}
  \and R.~P. Fender\inst{2}
  \and A. Marino\inst{3,4,5}
  \and P. Atri\inst{6,7}
  \and F. Carotenuto\inst{8}
  \and D. Williams-Baldwin\inst{9}
  \and M.~C. Baglio\inst{1}
  \and Jun-Ping Chen\inst{10,11}
  \and Wei Chen\inst{12}
  \and F. Coti Zelati\inst{3,13,1}
  \and F.~J. Cowie\inst{2}
  \and L. De Regt\inst{6,7}
  \and N. Grollimund\inst{14}
  \and A.~K. Hughes\inst{2}
  \and Yuan Liu\inst{12}
  \and I. Mariani\inst{1,15}
  \and N. Rea\inst{3,13}
  \and \SEM{L. Rhodes}\inst{16,17}
  \and P. Saikia\inst{18}
  \and C. Salvaggio\inst{1}
  \and K. Solomons\inst{19}
  \and J. van den Eijnden\inst{20}
  \and Kong-Jun Zhang\inst{21}
  \and Dan Zhu\inst{21}
  }

\authorrunning{Motta et al.}

\institute{%
Istituto Nazionale di Astrofisica, Osservatorio Astronomico di Brera, via E.\,Bianchi 46, 23807 Merate (LC), Italy\\
\email{sara.motta@inaf.it}
\and
University of Oxford, Department of Physics, Astrophysics, Denys Wilkinson Building, Keble Road, OX1 3RH, Oxford, United Kingdom
\and
Institute of Space Sciences (ICE, CSIC), Campus UAB, Carrer de Can Magrans s/n, E-08193 Barcelona, Spain
\and
Departamento de F\'isica, Universidad de Santiago de Chile (USACH), Av. V\'ictor Jara 3493, Estaci\'on Central, Chile
\and
INAF/IASF Palermo, via Ugo La Malfa 153, I-90146 Palermo, Italy
\and
ASTRON, Netherlands Institute for Radio Astronomy, Oude Hoogeveensedijk 4, 7991 PD Dwingeloo, The Netherlands
\and
Department of Astrophysics/IMAPP, Radboud University, P.O. Box 9010, 6500 GL Nijmegen, The Netherlands
\and
INAF -- Osservatorio Astronomico di Roma, Via Frascati 33, I-00078 Monte Porzio Catone (RM), Italy
\and
Jodrell Bank Centre for Astrophysics, School of Physics and Astronomy, The University of Manchester, Manchester M13 9PL, United Kingdom
\and
School of Physics and Astronomy, Sun Yat-Sen University, Zhuhai 519082, China
\and
CSST Science Center for the Guangdong-Hongkong-Macau Greater Bay Area, Sun Yat-Sen University, Zhuhai 519082, China
\and
National Astronomical Observatories, Chinese Academy of Sciences, Beijing 100101, China
\and
Institut d'Estudis Espacials de Catalunya (IEEC), 08860 Castelldefels (Barcelona), Spain
\and
Universit\'e Paris Cit\'e, Universit\'e Paris-Saclay, CEA, CNRS, AIM, F-91191 Gif-sur-Yvette, France
\and
Universit\`a degli Studi di Milano Bicocca, Dipartimento di Fisica, Piazza dell'Ateneo Nuovo 1, I-20126 Milano, Italy
\and
\SEM{Trottier Space Institute at McGill, 3550 Rue University, Montreal, Quebec H3A 2A7, Canada}
\and
\SEM{Department of Physics, McGill University, 3600 Rue University, Montreal, Quebec H3A 2T8, Canada}
\and
Department of Astronomy, Yale University, PO Box 208101, New Haven, CT 06520-8101, USA
\and
Department of Astronomy, University of Cape Town, Private Bag X3, Rondebosch 7701, South Africa
\and
Anton Pannekoek Institute for Astronomy, University of Amsterdam, Science Park 904, 1098 XH Amsterdam, The Netherlands
\and
South-Western Institute for Astronomy Research, Yunnan University, Kunming, Yunnan 650504, China
}

\date{Received ...; accepted ...}

\abstract
{After more than 25 years of near-Eddington activity and over six years in a Compton-thick obscured phase, the black-hole X-ray binary \grs\ has begun to fade in the radio to a level lower than ever previously recorded, in what may be the onset of its final descent towards quiescence. Given the current obscuration of the system, the direct emission from the accretion flow is not observable in the X-rays, and the radio emission, which is strongly coupled with the X-ray emission, is currently the best tracer of the accretion activity.}
{We aim to characterise the descent of \grs\ from its long-lived obscured radio plateau, and to establish whether the fading marks the end of the outburst that began in 1992 or a temporary switch-off.}
{We present the long-term MeerKAT monitoring of the source obtained as part of the ThunderKAT and X-KAT programmes, extending the record of \citet{Motta2021} to 2026 July, together with deep dual-band (L- and S-band) MeerKAT observations obtained through Director's Discretionary Time, and three \textit{Einstein Probe}/FXT X-ray pointings bracketing them.}
{Following a last Jy-level flare in early 2025, the 1.28\,GHz flux density declined monotonically to $\sim$0.2--0.4\,mJy by 2026 March, and after 2026 May 9 it was no longer formally detected in our 10-min monitoring observations. %
Deep MeerKAT observations in 2026 July, totalling $\sim$8\,hr on source per band, detect the source at $\sim$150\,\ujy\ in L band and $\sim$90\,\ujy\ in S band, the lowest radio flux densities recorded from the system since its discovery, corresponding to $L_{\rm R}\simeq2\times10^{28}$\,\ergs\ at 1.28\,GHz. The corresponding spectral index is $\alpha=-0.67\pm0.41$ across 1.28--2.62\,GHz, the first measured for this source in this regime.}
{The radio decline is the first direct evidence that the central engine itself is fading rather than merely receding further behind the absorber. The steep spectrum is consistent with either a combination of optically thick and thin emission, or an unusual, intrinsically steep core-jet spectrum. Should this mark the onset of quiescence, disc-instability models predict it to last of order $10^{4}$\,yr, so that the recent transition would not recur on any observationally accessible timescale. The switch-off of the central engine may shed light on the nature of the donor star in \grs, and on the origin of the very-high-energy gamma-ray emission in this system. }

\keywords{X-rays: binaries -- accretion, accretion disks -- black hole physics -- stars: jets -- radio continuum: stars -- stars: individual: \grs}

\maketitle

\section{Introduction}
\label{sec:intro}

In black-hole (BH) low-mass X-ray binaries (LMXBs), a stellar-mass black hole accretes matter via an accretion disc formed by gas stripped from a low-mass donor star. BH LMXBs are typically transient systems, alternating between long periods of quiescence, at luminosities of the order of $L\sim10^{30}$--$10^{33}$\,\ergs\ \citep[see][]{Wijnands2015}, and relatively short outbursts (lasting months, in most cases), during which their luminosity can reach and sometimes surpass the Eddington limit, $\sim10^{39}$\,\ergs. During outbursts these systems show clear repeating patterns of behaviour across various accretion states, each associated with mechanical feedback in the form of winds and relativistic jets \citep[e.g.][]{Fender2009,Ponti2012}. In the hard state the X-ray spectrum is dominated by Comptonisation in a hot flow close to the black hole, with the seed photons supplied either by a cool, geometrically thin accretion disc truncated at large radii or by synchrotron emission of the hot electrons themselves \citep[e.g.][]{PoutanenVeledina2014}. Alternatively, a significant fraction of the X-ray emission may originate at the base of a compact jet \citep[e.g.][]{Markoff2010}. \citet{Fender2001} established that black-hole X-ray binaries in their hard states typically show flat-spectrum radio emission ($\alpha\sim0$, where $S_\nu\propto\nu^{\alpha}$), which is regarded as the signature of partially self-absorbed compact jets, and is interpreted as synchrotron radiation from relativistic electrons in a magnetic field.

Across the hard and quiescent states of BH LMXBs a non-linear correlation is observed between the radio (\lr) and X-ray (\lx) luminosities, which provides a valuable tool to probe the connection between compact-jet production, traced by the radio emission, and the inner accretion flow or jet base, traced by the X-rays \citep{Hannikainen1998,Corbel2000,Corbel2003,Gallo2003,CrookMansour2026}. Whether quiescence is simply the low-luminosity tail of the hard state, with broadly similar spectral and physical properties, or a physically distinct regime in which compact jets need not always be produced, remains debated, as only a few systems have been detected in quiescence at radio wavelengths. 

Once the accretion rate drops below a critical value the structure of the accretion flow is expected to change, and several studies have reported a softening of the X-ray power-law spectral component (photon index $\Gamma\approx2.1$) relative to the canonical hard state \citep[e.g.][]{Corbel2006,Plotkin2013,Plotkin2015}. The origin of the quiescent X-ray emission, which is thought to arise from radiatively inefficient processes, likewise remains debated: it may be produced by the accretion flow, as in the advection-dominated accretion flow (ADAF) models \citep{Narayan1994,Narayan1995,Blandford1999}, or originate from synchrotron emission in the jet \citep{Markoff2001,Markoff2003,Markoff2005}, while a hybrid jet/ADAF model has also been proposed, which notably predicts jet-dominated X-ray emission in quiescence, leading to a steeper \lr--\lx\ correlation \citep{Yuan2005model,Yuan2005}. 

\grs\ is the archetypal Galactic microquasar, hosting a $12.4^{+2.0}_{-1.8}\,M_\odot$ black hole \citep{Reid2014} at a parallax distance of $9.4\pm0.7$\,kpc \citep{Reid2023}. It has the longest orbital period among the known XRB systems \citep[33.85\,d,][]{Steeghs2013}, and correspondingly the largest accretion disc of the class. Since its discovery at the onset of its 1992 outburst \citep{Castro-Tirado1992,Castro-Tirado1994} it sustained a near-Eddington $\sim$25-year-long activity phase, with recurrent compact jets and repeated superluminal ejections \citep{Mirabel1994,Fender1999,Rushton2010}. Following a 2018 transition to the hard state \citep{Negoro2018,Motta2021}, in mid-2019 the soft X-ray flux dropped significantly and the source became Compton-thick \citep[$N_{\rm H}\gtrsim10^{24}$\,cm$^{-2}$;][]{Miller2020,Koljonen2021,Motta2021,Athulya2023,Miller2025}, a state it has maintained ever since while remaining bright at radio wavelengths.

Since 2019 the X-ray flux has declined slowly. This decline was initially attributed to increasing intrinsic absorption rather than to a decreasing accretion rate, since the radio-bright jet remained clearly detected, at times flaring to Jy-level flux densities, indicating that the system was still highly active behind a thick local absorber \citep{Motta2021}. %
Recently, \grs\ was no longer formally detected by our routine MeerKAT monitoring. This prompted deep follow-up radio and X-ray observations in 2026 July, which recovered the system at the lowest radio flux density recorded since its discovery. 
In this paper we present the long-term radio and X-ray record of the decline and the deep dual-band MeerKAT observations obtained through Director's Discretionary Time (DDT), together with three \textit{Einstein Probe} pointings (Sects.~\ref{sec:obs} and \ref{sec:results_lc}--\ref{sec:results_ddt}). We then place the source on the radio/X-ray and luminosity--orbital-period planes (Sects.~\ref{sec:results_lrlx} and \ref{sec:results_porb}), discuss the nature of the fading (Sect.~\ref{sec:discussion}) and summarise our conclusions (Sect.~\ref{sec:conclusions}).

\section{Observations and data reduction}
\label{sec:obs}

\subsection{MeerKAT and AMI-LA long-term monitoring}
\label{sec:obs_monitoring}

We observed \grs\ (J2000 $19^{\rm h}15^{\rm m}11.56^{\rm s}$, $+10^\circ56'44.9''$) with the MeerKAT radio interferometer \citep{Jonas2016}, beginning on 2018 December 8, soon after the start of MeerKAT science operations. Observations were initially conducted as part of the ThunderKAT Large Survey Project \citep{Fender2016b}, and subsequently as part of the X-KAT Large Programme (PIs: Fender \& Motta), with an approximately weekly to monthly cadence. The data taken up to early 2021 were presented in \citet{Motta2021}. Additional data up to early 2023 were published in \citet{Sanchez-Sierras2023} and in De Regt et al. (submitted). Here we report data up to 2026 July. 

A typical monitoring observation consisted of 15 or 10 minutes\footnote{The ThunderKAT and X-KAT monitoring observations consisted of 15 min on source until 2025 October, and of 10 min on source from 2025 November onwards, at a small cost in sensitivity.} of on-source time, bookended by two 2-minute scans of the secondary (complex gain) calibrator, J2011$-$0644, with a 10-minute observation of the primary (flux and bandpass) calibrator, J1939$-$6342\footnote{All observations were pointed at a position slightly offset with respect to the nominal position of \grs, to avoid artefacts that can sometimes arise at the phase centre.}. The monitoring used the telescope's L-band receivers, at a central frequency of 1.28\,GHz across a 0.86\,GHz bandwidth (856--1712\,MHz). The correlator was configured to deliver either 4096 or 32768 channels, with an 8-second integration time per visibility point; in either case, data were binned down to 1024 channels for consistency before any further analysis. %

The data from the Arcminute Microkelvin Imager Large Array (AMI-LA; \citealt{Zwart2008,Hickish2018}), which observes at a central frequency of 15.5\,GHz, were largely presented in \citet{Motta2021} and were reduced and analysed following the same procedures described therein. This work includes all the data from early 2021 up to 2026 March 28, when the AMI-LA monitoring of \grs\ ended.  

\subsection{Deep dual-band DDT observations of 2026 July}
\label{sec:obs_ddt}

As part of a MeerKAT DDT request (proposal ID DDT-20260703-SM-01) we observed \grs\ for a total of $\sim$8\,hr on-source per band, in L-band and in S-band (central frequency 2.625\,GHz across a 0.88\,GHz bandwidth) in 2026 July. The data in each band were split into two separate observing blocks of $\sim$4\,hr on-source each: the L-band blocks were taken on 2026 July 7--8 and July 12--13, and the S-band blocks on 2026 July 8--9 and July 11--12, hence all four blocks were obtained within six days (see Table~\ref{tab:ddtlog}). Each block followed the same calibration set-up used for our monitoring, with the target scans interleaved with 2-minute scans of the secondary calibrator and bookended by scans of the primary calibrator. The correlator was configured to deliver 1024 channels, with an 8-second integration time per visibility point, and between 58 and 60 antennas were available (Table~\ref{tab:ddtlog}). 

\begin{table*}
\centering
\caption{Log of the dedicated observations of \grs\ used in this work: the deep MeerKAT DDT blocks (upper panel, Sect.~\ref{sec:obs_ddt}) and the \textit{Einstein Probe}/FXT pointings (lower panel, Sect.~\ref{sec:obs_ep}). Each MeerKAT block yielded $\sim$4\,hr on source, with a mean integration time of 8\,s per visibility point and 1024 channels across the band. The FXT observations were performed with both modules, A and B, in the readout modes listed, where FF and PW denote Full Frame and Partial Window, respectively. Times are UTC.}
\label{tab:ddtlog}
\begin{tabular}{lcccccc}
\hline\hline
\multicolumn{7}{c}{MeerKAT} \\
\hline
Block ID & Band & Start & End & Antennas & $\nu_{\rm c}$ & Bandwidth \\
 & & (UTC) & (UTC) & & (GHz) & (MHz) \\
\hline
1783450398 & L & 2026-07-07 18:55:15 & 2026-07-08 00:08:27 & 60 & 1.284 & 856 \\
1783536654 & S & 2026-07-08 18:52:49 & 2026-07-09 00:05:48 & 58 & 2.625 & 875 \\
1783802174 & S & 2026-07-11 20:37:52 & 2026-07-12 01:40:57 & 60 & 2.625 & 875 \\
1783891127 & L & 2026-07-12 21:19:14 & 2026-07-13 02:21:30 & 59 & 1.284 & 856 \\
\hline
\multicolumn{7}{c}{\textit{Einstein Probe}/FXT} \\
\hline
ObsID & Modules & Start & End & Mode & Exposure & Band \\
 & & (UTC) & (UTC) & (A / B) & (ks) & (keV) \\
\hline
06800001589 & A, B & 2026-06-27 09:51:36 & 2026-06-27 10:30:18 & FF / PW & 2.3 & 0.5--10 \\
06800001667 & A, B & 2026-07-29 19:39:53 & 2026-07-29 20:18:29 & FF / FF & 2.3 & 0.5--10 \\
06800001668 & A, B & 2026-08-01 06:33:08 & 2026-08-01 07:23:08 & FF / FF & 3.0 & 0.5--10 \\
\hline
\end{tabular}
\tablefoot{The MeerKAT start and end times bracket the full observing block, including the calibrator scans. For the FXT pointings the quoted exposure is the nominal duration of each observation, and the band is the nominal FXT passband; the flux upper limits of Sect.~\ref{sec:obs_ep} are instead quoted over 1--10\,keV. }
\end{table*}

\subsection{MeerKAT data reduction}
\label{sec:obs_reduction}

We reduced the data with a set of \textsc{python} scripts specifically tailored for the semi-automatic processing of MeerKAT data (\textsc{oxkat}\footnote{\url{https://github.com/IanHeywood/oxkat}}, \citealt{Heywood2020}), following the procedure described in \citet{Motta2021}. In brief, the flagging and the flux-scale, bandpass, delay and complex-gain calibration were carried out within \textsc{casa} \citep{CASAteam2022}, with additional flagging of the target field performed using the \textsc{tricolour} package\footnote{\url{https://github.com/ska-sa/tricolour/}}; each observation was imaged with \textsc{wsclean} \citep{Offringa2014}, using a deconvolution mask and a forced circular beam, and self-calibrated with \textsc{cubical} \citep{Kenyon2018}. The resulting images have typical resolutions of 4--8\,arcsec.

For the deep DDT observations, we first imaged the two 4-hr blocks of each band individually, as an internal check against source variability between the two epochs. 
Since the source is faint at these epochs, we measured its flux density with a
beam-constrained point-source fit, performed with \textsc{casa} \texttt{imfit}.
For each image we fitted a single two-dimensional Gaussian in the image plane, holding
the Gaussian shape fixed to the restoring beam of that image, so that the only
free parameters describing the source are its amplitude and its centroid.
The fit was performed within a circular region of radius four times the beam FWHM, centred on the nominal source position.
A constant zero level was fitted simultaneously with the Gaussian to account for any unmodelled constant underlying flux. The fitted zero levels are comparable to or slightly higher than the local rms (Table~\ref{tab:imfit}).
The local rms was measured separately, as the median absolute deviation within an annulus of inner and outer radii $40''$ and $120''$ centred on the source, scaled to a Gaussian-equivalent standard deviation, and is quoted alongside each
measurement in Table~\ref{tab:imfit}. We regard the source as detected where
the fitted amplitude exceeds three times that local rms, and quote a
$3\sigma$ upper limit otherwise. The uncertainties on the flux densities are the formal errors returned by the fit, floored at the local rms. When a spectral index is derived across the two
bands we add a flux-scale uncertainty of 10\% per band in quadrature, treating
the bands as independent. This flux-scale uncertainty is larger than the 5\% adopted for the monitoring observations, to allow for residual calibration errors in the long, stacked exposures, as well as for any intrinsic variability of the source between the blocks.%
The results of the fits to the individual blocks are collected in Table~\ref{tab:imfit}.

Because the four blocks were taken as two consecutive-night pairs, each comprising one L-band and one S-band block, they also afford two quasi-simultaneous measurements of the radio
spectrum. We therefore paired each L-band block with the S-band block closest
in time, the two blocks of each pair being separated by about one day, and computed
for each pair the two-point spectral index between the two image frequencies.
The uncertainty on
$\alpha$ has a statistical part, propagated from the formal fit errors, and a
systematic part arising from the 10\% per-band flux-scale uncertainty
described above.
When the two pairs are combined we weight them by their statistical errors alone and add the flux-scale term afterwards. 

Finally, we stacked the two (self-)calibrated measurement sets of each band in the visibility domain, imaging them jointly with \textsc{wsclean} to produce a single deep image per band. These were fitted in exactly the same way, and the results are given in the last two rows of Table~\ref{tab:imfit}. The flux densities agree closely with those of the contributing blocks.

\subsection{\textit{Einstein Probe} X-ray observations}
\label{sec:obs_ep}
To assess the X-ray status of the source we obtained three observations (of 2.3, 2.3 and 3.0 ks exposures, respectively) with the Follow-up X-ray Telescope (FXT) on board \textit{Einstein Probe} \citep{Yuan2025} on 2026 June 27, July 29 and August 1 (ObsIDs: 06800001589, 06800001667 and 06800001668). The first observation was performed with module A in Full Frame (FF) mode and module B in Partial Window (PW) mode, while in the remaining two both modules were operated in FF mode. The data were processed with the \texttt{fxtchain} pipeline in the FXT Data Analysis Software Package (FXTDAS v.1.30). Circular regions of $60\arcsec$ radius, centred on the coordinates of the source and on a nearby, source-free area, were then adopted to extract the source and background spectra, respectively. No X-ray source is significantly detected within 1\,arcmin of the source coordinates in any of the observations, as reported by \citet{Marino2026_atel}.
In order to estimate $3\sigma$ upper limits on the X-ray fluxes for each observation, we adopted two complementary methods. First, using the \texttt{sosta} (source statistics) tool within \textsc{ximage}, we measured the number of events within the selected source region and estimated an upper limit on the count rate, accounting for vignetting, exposure, the point-spread function, and background counts. We then converted the count-rate upper limit into an X-ray flux by simulating source spectra (see below) using a fixed spectral model with variable normalisation, and iteratively varying the normalisation until the simulated count rate matched the upper limit. The 1--10\,keV flux of the resulting spectrum, calculated by including the \texttt{cflux} component in the model, was then adopted as the corresponding upper limit on the flux. We assumed a fiducial spectral shape consisting of an absorbed power-law model, with $N_{\rm H}$ fixed to the average value in the direction of the source, as determined using the \texttt{nH} tool\footnote{\url{https://heasarc.gsfc.nasa.gov/cgi-bin/Tools/w3nh/w3nh.pl}; see \citet{HI4PICollaboration} for further details}, i.e., $1.4\times10^{22}$ cm$^{-2}$, standard abundances \citep{Wilms2000}, and $\Gamma=1.5$, while leaving the normalisation $K_{\rm powerlaw}$ free. As a second, independent, approach, we directly modelled the source spectra with \textsc{xspec} \citep{Arnaud1996}, after grouping them with the FTOOLS command \texttt{ftgrouppha} to a minimum of one count per bin, and fitting them with the Cash statistic \citep{Cash1979}. We adopted the same absorbed power-law model, again leaving $K_{\rm powerlaw}$ as the only free parameter. Following the fit, we derived an upper limit on $K_{\rm powerlaw}$ and used the \texttt{cflux} component to calculate the corresponding 1--10\,keV flux. The two methods yielded consistent flux upper limits for the individual observations, the most constraining being $8.8\times10^{-14}$\,erg\,cm$^{-2}$\,s$^{-1}$ for ObsID 06800001668. By then assuming that the spectral shape of the source did not change throughout the three executed FXT observations, we modelled the six spectra together, i.e., one spectrum per module (2) per observation (3), to obtain a deeper upper limit of $8.1\times10^{-14}$\,erg\,cm$^{-2}$\,s$^{-1}$, corresponding to an observed luminosity of $L_{\rm X}\lesssim8.6\times10^{32}$\,\ergs\ at 9.4\,kpc. We note that in the last few years, \grs\ has been in a very obscured state, with intrinsic absorption exceeding the one expected in the direction of the source even by a few orders of magnitude \citep[e.g.][]{Motta2021}. Because of the X-ray non-detection, we cannot assess whether the source is still affected by significant obscuration, preventing us from using a reliable value for the intrinsic $N_{\rm H}$ in our upper-limit estimation procedure. Because of the above, we caution that the upper limit on the X-ray luminosity measured here might be underestimated and has to be considered as tentative.

\section{Results}
\label{sec:results}

\subsection{The X-ray and radio dimming}
\label{sec:results_lc}

\grs\ disappeared from the X-ray sky gradually, and at a pace that differed between the hard and the soft band. 
To quantify this, we rebinned the \textit{Swift}/BAT (15--50\,keV; \citealt{Krimm2013}) and MAXI/GSC (2--12\,keV; \citealt{Matsuoka2009,Mihara2011}) daily light curves into inverse-variance weighted 7-day averages, and we adopted a signal-to-noise ratio of 3 as the detection threshold, which corresponds to empirical $3\sigma$ sensitivities of $\sim$18 and $\sim$5\,mCrab per weekly bin for BAT and MAXI, respectively, and $\sim$9 and $\sim$2\,mCrab in 30-day averages. %

Stacking the MAXI data in half-year blocks, we find that the rate declines until mid-2025 and then settles onto a constant floor of $(10.1\pm1.0)\times10^{-3}$\,ph\,cm$^{-2}$\,s$^{-1}$ ($\sim$2.7\,mCrab), whereas the BAT rate over the same interval stacks to zero to within its errors ($(0.1\pm3.4)\times10^{-4}$\,ct\,cm$^{-2}$\,s$^{-1}$). A residual soft signal of this level is incompatible with the deep \textit{NuSTAR} non-detection of 2025 November 17 \citep{Gandhi2026}, and we therefore attribute it to source confusion or a background systematic in this crowded Galactic-plane field, to which MAXI/GSC may be vulnerable while BAT is not. We subtract this floor from the MAXI series, propagating its uncertainty, and we assume it to be constant in time.

\begin{figure*}
\centering
\includegraphics[width=0.93\linewidth]{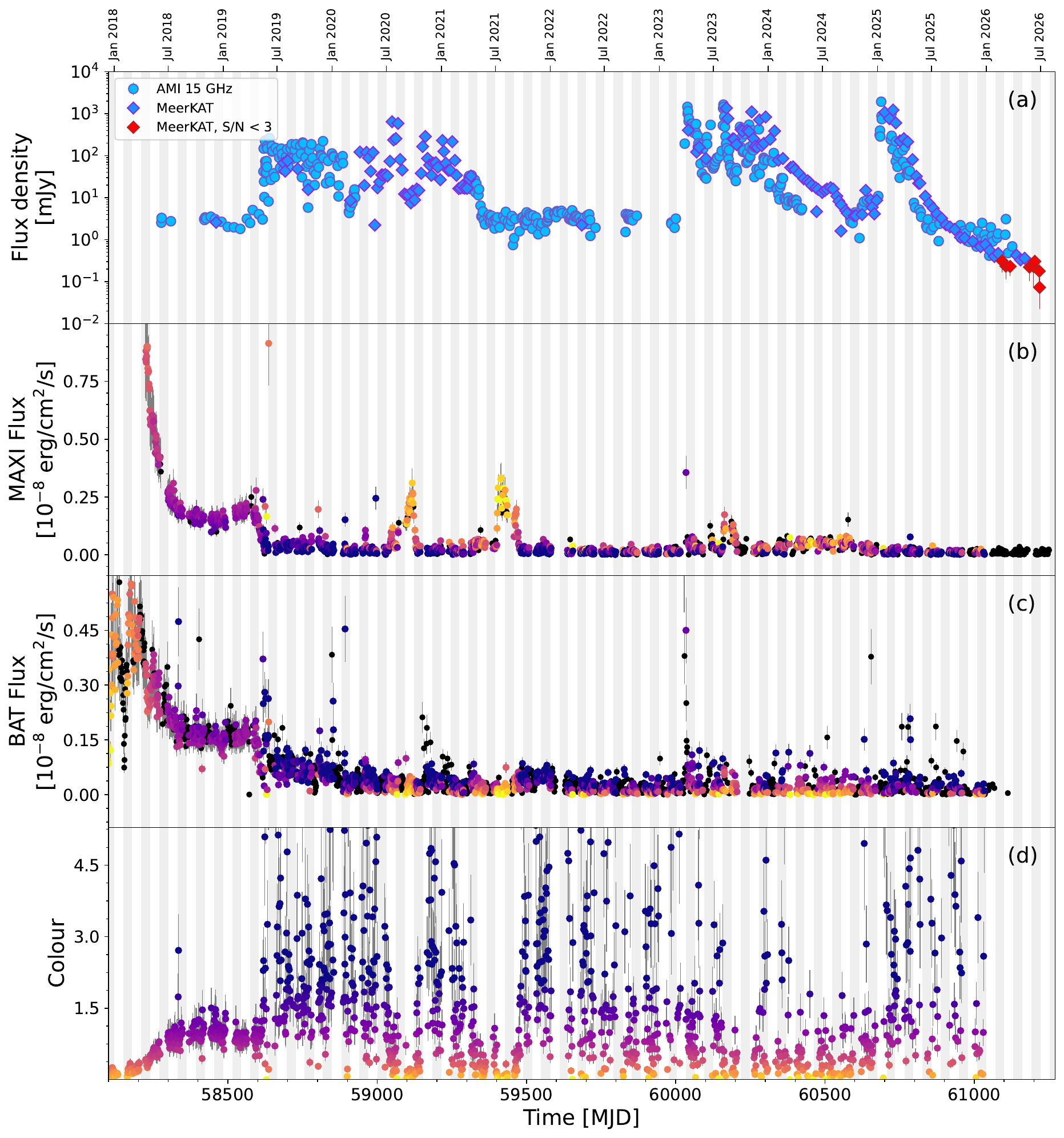}
\caption{Long-term radio and X-ray light curves of \grs\ from 2018 January to 2026 July. (\textit{a}) AMI-LA 15.5\,GHz (circles) and MeerKAT 1.28\,GHz (diamonds) flux densities, with red diamonds marking MeerKAT epochs below the $3\sigma$ detection threshold (Sect.~\ref{sec:results_lc}). (\textit{b}) MAXI/GSC 2--12\,keV fluxes, not corrected for the constant floor discussed in Sect.~\ref{sec:results_lc}, and (\textit{c}) \textit{Swift}/BAT 15--50\,keV fluxes. (\textit{d}) X-ray colour, defined as the ratio of the BAT to the MAXI flux. The points in panels \textit{b}--\textit{d} are colour-coded by the X-ray colour of panel \textit{d}, with warmer colours corresponding to softer spectra, while black points mark epochs for which the colour could not be computed. The vertical scales of panels \textit{b} and \textit{c} are chosen to display the evolution after the 2019 transition, so that the brightest points of early 2018 fall off scale. See Figs.~2 and 3 of \citet{Motta2021} for details. In each panel, white and grey bands mark calendar months.}
\label{fig:lc}
\end{figure*}

BAT detected the source in nearly all weekly bins up to the end of 2020. The detected fraction fell to 66\% in 2021, and the first blind stretch longer than a month appeared in 2021 May--August. The detection fraction then declined steadily through 55\% (2022) and 21\% (2023) to 6\% in 2024. A run of weekly BAT detections between 2025 January and May, likely associated with renewed hard flaring, briefly interrupted the fade, and the last hard X-ray detection was recorded on 2025 September 19 (MJD 60937). 
MAXI/GSC detected 82\% of the 2024 weekly bins, during a period of renewed soft X-ray activity in 2023--2024 (Fig.~\ref{fig:lc}, panel \textit{b}), with the last weekly MAXI detection on 2025 April 18 (MJD 60783) coinciding with the same flaring episode seen by BAT, followed by an uninterrupted blind stretch of $\sim$460 days, extending to the end of the available data (2026 July).
Hence \grs\ has been persistently undetected by the all-sky monitors in the soft X-rays since 2025 April, at the end of the last flaring episode (2025 January--May), and in the hard X-rays since 2025 September, rather than progressively through 2026 as the uncorrected MAXI light curve would suggest. Since \grs\ is presumably still Compton-thick, these are absorbed fluxes, and the non-detections therefore do not by themselves distinguish a fall in the intrinsic luminosity from a further rise in the local obscuration.

The radio behaviour was markedly different, in that the source remained formally detected at the $3\sigma$ level at every pointed monitoring epoch until 2026 February, that is, for several months after the X-rays had vanished, and it was still detected, albeit marginally, as late as 2026 May. Throughout this section we apply to the monitoring the detection criterion of Sect.~\ref{sec:obs_reduction}, that is, a fitted amplitude exceeding three times the local image rms, and we quote total uncertainties, which combine the local rms with a 5\% flux-scale systematic. The monitoring flux densities were measured with the same beam-constrained fit, including a constant zero level fitted simultaneously with the Gaussian.
MeerKAT detected \grs\ at all but eight of the 155 L-band epochs obtained between 2018 December and 2026 June, and all eight sub-threshold epochs belong to the final decline of 2026. These are the epochs of 2026 February 22, March 7, March 21, May 24, June 7, June 11, June 26 and June 27, with fitted amplitudes between $73$ and $313$\,\ujy\ at $1.4$ to $2.96$ times the local rms. The three intervening epochs of 2026 April 12, April 25 and May 9 are formal detections, but only marginally so, at $3.4$, $3.6$ and $3.1$ times the local rms respectively, so that the transition from detection to non-detection is gradual and is set as much by the epoch-to-epoch variation in image noise as by the fading of the source itself. 

The last formal detection was obtained on 2026 May 9, at $360\pm120$\,\ujy. From the last major flare, which peaked above 1\,Jy in 2025 January--February and had decayed to $\sim$250\,mJy by 2025 April, the 1.28\,GHz flux density declined to $\sim$1.1\,mJy by 2025 October and $\sim$0.55\,mJy by 2026 January, and then hovered between $\sim$0.2 and $\sim$0.4\,mJy from late 2026 February onwards in the fortnightly monitoring runs, at marginally significant levels. 
After 2026 May 9, none of the subsequent L-band epochs (May 24 and the four epochs of June) reaches our detection threshold in the 10-min runs, although a fitted amplitude of $305\pm104$\,\ujy\ on June 11 falls only marginally short of the threshold, at $2.96$ times the local rms. A dedicated, longer (15-min) S-band observation obtained within the X-KAT monitoring on June 7 yielded a sub-threshold amplitude of $140\pm49$\,\ujy, i.e. 2.85 times the local rms. The preliminary non-detections reported in \citet{Motta2026} hence correspond to a drop below the sensitivity of the X-KAT monitoring observations.

The AMI-LA 15.5\,GHz monitoring continued to detect the source in 29 of its 34 epochs between 2025 October and 2026 March, at a median of $\sim$1.1\,mJy (range 0.4--3.1\,mJy), with its last detection on 2026 March 28 ($0.69\pm0.14$\,mJy), which is also the last AMI-LA observation of \grs. Quasi-simultaneous MeerKAT/AMI pairs over this interval imply flat to inverted spectra ($\alpha\approx0$ to $+1$), suggesting that low-level, self-absorbed flaring continued until at least 2026 March. 

\subsection{Deep dual-band measurements}
\label{sec:results_ddt}
The deep DDT radio observations obtained 10--16\,d after the last monitoring epoch of 2026 June 27 (Sect.~\ref{sec:obs_ddt}) recover the source at a level compatible with the upper limits from the monitoring, i.e., $\sim$150\,\ujy\ in L band and $\sim$90\,\ujy\ in S band, which supports the view that by late 2026 June the source had fallen below the sensitivity of the routine monitoring while remaining present. These are the lowest radio flux densities recorded from \grs\ since its discovery in 1992, roughly an order of magnitude below the $\sim$1\,mJy level reached in late 2025.

The results of the beam-constrained point-source fits described in
Sect.~\ref{sec:obs_reduction} are collected in Table~\ref{tab:imfit}, for the
individual observing blocks and for the two stacked images, one per band.
The source is formally detected in all four individual blocks. %
The two L-band measurements, taken five days apart, agree to $0.1\sigma$, and the two S-band measurements, taken three days apart, to $0.2\sigma$, hence we find no evidence for variability over six days and the four blocks may be treated as a single dual frequency epoch. The stacked images, at $150\pm34$\,\ujy\ in L band and $93\pm11$\,\ujy\ in S band, agree with the blocks that went into them to $0.003$ and $0.4\sigma$ respectively.

The two consecutive-night pairs yield $\alpha=-0.70\pm0.36$ and $\alpha=-0.78\pm0.32$ across 1.28--2.62\,GHz, the first spectral measurements obtained for this source in this regime. The two values agree with each other and also fully agree with the in-band $\alpha$ calculated both in the individual S-band and L-band blocks. Combining the values from the two pairs we obtain  $\alpha=-0.74\pm0.24\,({\rm stat})\pm0.20\,({\rm sys})$ ($-0.74\pm0.31$), while the $\alpha$ from the stacked images gives $\alpha=-0.67\pm0.36\,({\rm stat})\pm0.20\,({\rm sys})$ ($-0.67\pm0.41$ overall), consistent with the pair combination. We adopt the stacked value as the reference spectral index in what follows.

\begin{table*}
\centering
\caption{Two-dimensional Gaussian fits to the deep MeerKAT DDT images of \grs, and the resulting two-point spectral indices. All fits were performed with \textsc{casa} \texttt{imfit} on primary-beam-corrected images, following the same procedure as the rest of our MeerKAT monitoring. The blocks are grouped into the two consecutive-night L/S pairs from which $\alpha$ was computed, and are followed by the two stacked images. Radio luminosities are quoted at the image frequency of each row and require no spectral assumption. The uncertainties on the flux densities are the formal fit errors, and those on $\alpha$ are statistical; every value of $\alpha$ carries in addition a systematic uncertainty of $\pm0.20$, arising from an adopted flux-scale uncertainty of 10\% per band.}
\label{tab:imfit}
\small
\setlength{\tabcolsep}{5pt}
\begin{tabular}{llcccccccc}
\hline\hline
Pair & Block ID & Band & MJD & Beam & rms & $b$ & $S_\nu$ & $L_{\rm R}$ & $\alpha$ \\
 & & & & ($''$) & (\ujb) & (\ujb) & (\ujy) & ($10^{28}$\,\ergs) & \\
\hline
\multirow{2}{*}{P1} & 1783450398 & L & 61228.897 & 6.91 & 32  & 53.4 & $148\pm32$   & $2.0\pm0.4$ & \multirow{2}{*}{$-0.70\pm0.36$} \\
                    & 1783536654 & S & 61229.895 & 4.07 & 12  & 57.9 & $90\pm12$    & $2.5\pm0.3$ & \\
\multirow{2}{*}{P2} & 1783802174 & S & 61232.965 & 4.31 & 9.5 & 33.0 & $86.8\pm9.5$ & $2.4\pm0.3$ & \multirow{2}{*}{$-0.78\pm0.32$} \\
                    & 1783891127 & L & 61233.993 & 7.70 & 30  & 37.8 & $152\pm30$   & $2.1\pm0.4$ & \\
\hline
\multirow{2}{*}{Stacks} & L stack$^{\rm a}$ & L & 61231.445 & 7.22 & 34 & 15.1 & $150\pm34$ & $2.0\pm0.5$ & \multirow{2}{*}{$-0.67\pm0.36$} \\
                        & S stack$^{\rm b}$ & S & 61231.430 & 4.17 & 11 & 55.6 & $93\pm11$  & $2.6\pm0.3$ & \\
\hline
\end{tabular}
\tablefoot{MJD refers to the mid-point of each observing block. The beam column gives the FWHM of the circular Gaussian to which each fit was constrained, and $b$ is the constant zero level fitted simultaneously with the Gaussian. %
The radio luminosity is $L_{\rm R}=4\pi d^{2}\nu S_\nu$ evaluated at the image frequency of each row, for $d=9.4$\,kpc, so that it involves no spectral assumption. %
We note that the S-band entries are therefore monochromatic luminosities at 2.62\,GHz, and are not directly comparable with the 1.28\,GHz values used on the radio/X-ray plane (Sect.~\ref{sec:results_lrlx}), where the S-band point is instead shifted to 1.28\,GHz under the flat-spectrum assumption. %
$^{\rm a}$ Blocks 1783450398 and 1783891127, stacked in the visibility domain. $^{\rm b}$ Blocks 1783536654 and 1783802174, stacked in the visibility domain. For the stacks, MJD is the mean of the mid-points of the two contributing blocks.
}
\end{table*}

\subsection{The radio/X-ray plane}
\label{sec:results_lrlx}

\begin{figure*}
\centering
\includegraphics[width=0.93\linewidth]{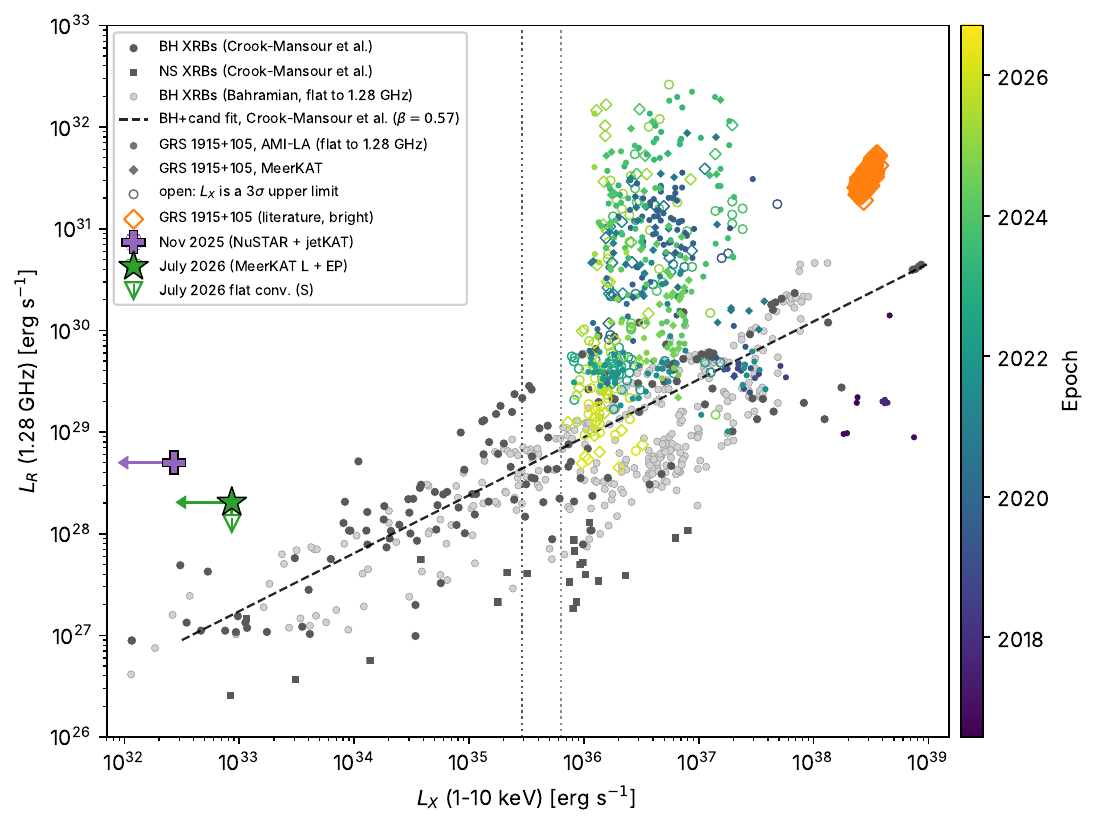}
\caption{Radio/X-ray plane at the native MeerKAT L-band frequency. Dark grey symbols are the hard- and quiescent-state ThunderKAT sample of \citet{CrookMansour2026} (circles: black holes and candidates, squares: neutron stars), and the dashed line is their correlation for black holes and candidates. Light grey circles are the black holes of \citet{Bahramian2018}, and open orange diamonds the historical bright-state values of \grs, both shifted from 5 to 1.28\,GHz assuming a flat spectrum, as are all flat-convention points of \grs. Coloured symbols are the 525 quasi-simultaneous pairs of \grs\ with a radio detection, colour-coded by epoch: circles for AMI-LA (15.5\,GHz, flat-shifted) and small diamonds for MeerKAT, filled for X-ray detections and open for MAXI $3\sigma$ upper limits on \lx\ (the 18 radio upper limits are not shown). Vertical dotted lines are the stacked $3\sigma$ MAXI and BAT depths after their last detections. The purple cross is the 2025 November \textit{NuSTAR} upper limit (1--10\,keV, $\Gamma=1.25$) paired with the jetKAT S-band detection of 2025 December 2. The green star is the 2026 July MeerKAT L-band measurement ($L_{\rm R}=(2.0\pm0.5)\times10^{28}$\,\ergs) at the joint \textit{Einstein Probe} upper limit (Sect.~\ref{sec:obs_ep}), and the open green triangle is the flat-convention S-band stack. \lx\ is the observed, absorbed luminosity throughout.}
\label{fig:lrlx}
\end{figure*}

We constructed the radio/X-ray plane to determine the position of \grs\ with respect to the overall binary population. The comparison sample (dark grey points in Fig.~\ref{fig:lrlx}) is the MeerKAT L-band compilation of \citet{CrookMansour2026}, restricted to the hard and quiescent states, together with the correlation fit to BHs and BH candidates, $\log(L_{\rm R}/L_{\rm R,0}) = \xi + \beta \log(L_{\rm X}/L_{\rm X,0})$, with $\beta=0.57^{+0.03}_{-0.03}$, $\xi=0.22^{+0.04}_{-0.04}$, an intrinsic scatter $\sigma_\epsilon=0.48^{+0.04}_{-0.03}$\,dex, and pivots $L_{\rm R,0}=4.54\times10^{28}$ and $L_{\rm X,0}=7.51\times10^{35}$\,\ergs. For comparison we also show the black-hole sample of \citet{Bahramian2018} (light grey), together with the historical bright-state measurements of \grs\ itself from that compilation, previously shown by \citet{Motta2021}. Throughout, radio luminosities are computed as $L_{\rm R}=4\pi d^{2}\nu S_\nu$ at the native MeerKAT frequency, $\nu=1.28$\,GHz, which is the natural choice both for a MeerKAT-based comparison sample and for our own MeerKAT data.  Quantities measured at other frequencies (the AMI monitoring, the MeerKAT S-band core measurement of 2025 December 2 described below, and the 5\,GHz values of the \citet{Bahramian2018} compilation) are scaled to 1.28\,GHz assuming a flat spectrum ($\alpha=0$). X-ray luminosities are observed (not corrected for the local absorption) 1--10\,keV values, and we adopt $d=9.4$\,kpc for \grs.

To trace the evolution of \grs\ across the plane, regardless of the accretion state of the system,\footnote{The radio/X-ray plane is built from hard-state X-ray and radio fluxes, but for diagnostic purposes we show all the available radio and X-ray data on the plane.} we matched every epoch of the MeerKAT and AMI monitoring to the floor-corrected MAXI light curve (Sect.~\ref{sec:results_lc}). 
The X-ray rate at each radio epoch is taken from the same-day MAXI bin where that bin is itself a $3\sigma$ detection. Failing that, we stack the
daily bins falling within $\pm3$\,d of the radio epoch into an
inverse-variance weighted mean,\footnote{We used $\bar r = \sum_i w_i r_i / \sum_i w_i$ where $r_i$ are the MAXI rates, $w_i = \sigma_i^{-2}$ and
$\sigma_{\bar r} = (\sum_i w_i)^{-1/2}$.} and we widen the window to
$\pm7$\,d if the narrower stack does not reach $3\sigma$. Where no window reaches $3\sigma$ we assign the $3\sigma$ upper limit of the $\pm7$\,d
stack, and radio epochs without MAXI coverage within $\pm7$\,d are
excluded altogether. We note that this stacking carries no weighting by
temporal proximity, so that a bin three days from the radio epoch
contributes as much as a same-day bin of equal uncertainty, and it
therefore assumes the X-ray flux to be constant across the window. %
Of the 543 pairs, 122 rest on a same-day detection, 145 on the $\pm3$\,d
stack and 86 on the $\pm7$\,d stack, while the remaining 190 carry
$3\sigma$ upper limits on \lx.

We stress that the flat-spectrum placement is only approximate for the subset of our data taken at other frequencies: the quasi-simultaneous MeerKAT/AMI pairs imply a median spectral index of $\alpha\simeq-0.4$ between 1.28 and 15.5\,GHz, expected for a flaring source, hence the AMI points (circles in Fig.~\ref{fig:lrlx}), which carry the flat assumption across the largest frequency ratio, likely underestimate the 1.28\,GHz luminosity, by a factor of $\sim$3 for the median spectral index.
The MeerKAT points, being native to the frequency of the plane, carry no such uncertainty, and the deep DDT L-band measurement of 2026 July in particular requires no spectral extrapolation at all (see below).

Throughout the obscured phase the source sits far above the standard black-hole track, by up to three orders of magnitude in $L_{\rm R}$ at a given observed $L_{\rm X}$, which is the expected signature of an unabsorbed radio flux paired with a heavily absorbed X-ray flux. In this regard, we note that the X-ray luminosities are not corrected for absorption as there is no way to accurately constrain the column density affecting the X-ray fluxes. The intrinsic \lx\ may be substantially higher and the true positions may lie well to the right of those plotted, likely close to the historical values (orange points) in the figure. 
The vertical dotted lines in Fig.~\ref{fig:lrlx} mark the $3\sigma$ depths reached by stacking all MAXI and BAT data after their respective last detections, $L_{\rm X}<2.9\times10^{35}$ and $<6.3\times10^{35}$\,\ergs\ (the latter converted from 15--50\,keV assuming $\Gamma=1.8$). %

Two deep constraints can be added to the radio/X-ray plane. All the X-ray limits used in this work are collected in Table~\ref{tab:xray}.
The first constraint comes from \textit{NuSTAR}. On 2025 November 17 a 32\,ks \textit{NuSTAR} observation did not detect the source, setting an upper limit to the luminosity at $L_{3-78\,{\rm keV}}<1.4\times10^{33}$\,\ergs\ \citep{Gandhi2026}, which translates into a 1--10\,keV X-ray luminosity of $L_{\rm X}<2.7\times10^{32}$\,\ergs, assuming $\Gamma=1.25$.
Pairing this limit with a radio measurement requires some care, since the \textit{NuSTAR} pointing fell during the decline of a major radio flare, a phase in which the emission is dominated by the cooling of previously ejected material rather than by the compact core. The epochs closest in time do not separate the two components: our L-band monitoring detected the source at $918\pm121$\,\ujy\ some 19\,hr before the \textit{NuSTAR} pointing, and AMI provides a non-detection three days after it, with a $3\sigma$ limit of 0.43\,mJy ($L_{\rm R}<5.8\times10^{28}$\,\ergs\ at 1.28\,GHz under the flat-spectrum assumption), but in both cases the measured flux density is the sum of core and ejecta. Since the radio/X-ray correlation is calibrated on the core emission of compact jets, an unresolved flux density measured during the decay of an ejection event would place the source spuriously high on the plane.
We therefore pair the \textit{NuSTAR} limit with a deeper 1~hr MeerKAT S-band observation taken on 2025 December 2, 15 days after the X-ray pointing, as part of the jetKAT programme (PI: Motta, Solomons et al. in prep.). At S-band the angular resolution is sufficient to separate the core from the resolved jet components, hence we can isolate the core contribution. The core is detected at $368\pm37$\,\ujy, corresponding to $L_{\rm R}=(5.0\pm0.5)\times10^{28}$\,\ergs\ at 1.28\,GHz under the flat-spectrum assumption, and consistent with the shallower AMI limit. Assuming that the core remained approximately steady over these 15 days, we accept the longer interval in exchange for a measurement of the core alone.

The second constraint is provided by the deep MeerKAT observations of 2026 July (Sect.~\ref{sec:results_ddt}) and the \textit{Einstein Probe}. The stacked L-band flux density of $150\pm34$\,\ujy\ corresponds to $L_{\rm R}=(2.0\pm0.5)\times10^{28}$\,\ergs. The S-band stack, placed at 1.28\,GHz under the flat-spectrum assumption, is shown for comparison (open triangle). The measured spectral index between the two bands is $\alpha=-0.67\pm0.41$ (Sect.~\ref{sec:results_ddt}), so the flat placement of the S-band point should be regarded as indicative only. We pair this point with the joint \textit{Einstein Probe} FXT limit of $L_{\rm X}\lesssim8.6\times10^{32}$\,\ergs\ (Sect.~\ref{sec:obs_ep}).

Note that the pairing is not strictly simultaneous, as the three FXT pointings bracket the radio blocks, the nearest preceding the first block by 10\,d and following the last block by 16\,d. However, the stability of the radio flux across the six-day campaign suggests that the source is unlikely to have varied substantially over this interval. 

\begin{table*}
\centering
\caption{X-ray luminosity constraints on \grs. All values are $3\sigma$ upper limits on the luminosity at $d=9.4$\,kpc. The observed luminosities are quoted both in the native band of each instrument and converted to 1--10\,keV as described in Sects.~\ref{sec:obs_ep} and \ref{sec:results_lrlx}. The corrected luminosities are the intrinsic 1--10\,keV luminosities obtained by correcting the native-band limit for transmission through an effective local column density $N_{\rm H}=1.7\times10^{24}$\,cm$^{-2}$ (Sect.~\ref{sec:disc_obscuration}), accounting for photoelectric absorption and Compton scattering only.}
\label{tab:xray}
\begin{tabular}{llccccc}
\hline\hline
Instrument & Interval & Exposure & Band & $L_{\rm band}^{\rm obs}$ & $L_{1-10}^{\rm obs}$ & $L_{1-10}^{\rm corr}$ \\
 & & & (keV) & (\ergs) & (\ergs) & (\ergs) \\
\hline
MAXI/GSC$^{\rm a}$ & 2025 Apr 21 -- 2026 Jul 31 & 259\,d & 2--12 & $<2.1\times10^{35}$ & $<2.9\times10^{35}$ & $<2.0\times10^{37}$ \\
\textit{Swift}/BAT$^{\rm a}$ & 2025 Jul 6 -- 2026 Mar 13 & 101\,d & 15--50 & $<5.0\times10^{35}$ & $<6.3\times10^{35}$ & $<4.7\times10^{36}$ \\
\textit{NuSTAR}$^{\rm b}$ & 2025 Nov 17 & 32\,ks & 3--78 & $<1.4\times10^{33}$ & $<2.7\times10^{32}$ & $<7.9\times10^{33}$ \\
\textit{Einstein Probe}/FXT$^{\rm c}$ & 2026 Jun 27 -- Aug 1 & 7.6\,ks & 1--10 & $<8.6\times10^{32}$ & $<8.6\times10^{32}$ & $<1.7\times10^{35}$ \\
\hline
\end{tabular}
\tablefoot{$^{\rm a}$ Inverse-variance weighted stack of all daily bins after the last daily detection (Sect.~\ref{sec:results_lc}); the MAXI rates are corrected for the constant floor. $^{\rm b}$ From \citet{Gandhi2026}. $^{\rm c}$ Joint fit of the three FXT pointings of Table~\ref{tab:ddtlog} (Sect.~\ref{sec:obs_ep}).}
\end{table*}

\subsection{Quiescent luminosities versus orbital period}
\label{sec:results_porb}

In quiescence, both the X-ray and the radio luminosity of BH LMXBs are expected to increase with the orbital period, essentially because the mass-transfer rate and the disc size, and with them the rate at which matter reaches the inner hot flow, grow with the binary separation (\citealt{Menou1999,Garcia2001}). To place the 2026 July dimming of \grs\ in this context, we assembled a sample comprising all transient BH LMXBs, dynamically confirmed or strong candidates, with a measured orbital period and a published quiescent X-ray measurement or upper limit (Table~\ref{tab:porbsample}). The sample counts 19 such systems with X-ray constraints (16 detections and 3 upper limits), of which 17 also carry radio constraints (5 detections and 12 upper limits). Radio luminosities were rescaled homogeneously from the published flux densities (mostly observed at 5--8.5\,GHz) to the 1.28\,GHz convention of Sect.~\ref{sec:results_lrlx}, assuming a flat spectrum. The X-ray luminosities are as published in bands that differ between references (0.5--10 to 3--9\,keV): for the soft power-law spectra of quiescent systems ($\Gamma\approx2$) the corresponding band conversions change the luminosities by at most a factor of $\sim$2 (reached only for the narrow 3--9\,keV band, and below $\sim$1.6 otherwise), smaller than the factor-of-a-few variability that quiescent systems display in both bands. For this reason, we do not correct for the different energy bands, assuming that the intrinsic scatter of the points dominates over the uncertainties of the individual measurements. The distances adopted to compute the luminosities are listed in Table~\ref{tab:porbsample}. Four additional systems have measured periods, but lack published quiescent luminosities, and are listed in the table for completeness.

Given the substantial fraction of upper limits, we assess the significance of each correlation with a generalised Kendall rank test \citep{Isobe1986} that uses only pairs whose ordering is determinate in the presence of upper limits, with the one-sided $p$-value drawn from $5\times10^{3}$ random permutations, and we model each relation as $\log_{10} L = a + b\,\log_{10}(P_{\rm orb}/10\,{\rm hr})$ with a censored (Tobit) maximum-likelihood fit, in which detections contribute Gaussian densities and upper limits the corresponding cumulative probabilities, and where the dispersion $\sigma$ absorbs measurement errors and intrinsic scatter alike. Parameter uncertainties are the 16--84\% intervals from $10^{4}$ bootstrap resamples of the systems. \grs\ is excluded from all fits, and its position is compared a posteriori with the relation defined by the rest of the population; we represent it by the native 1.28\,GHz L-band measurement of Sect.~\ref{sec:results_lrlx}, which requires no spectral assumption, and by the absorbed joint \textit{Einstein Probe} limit.

\begin{figure}
\centering
\includegraphics[width=\columnwidth]{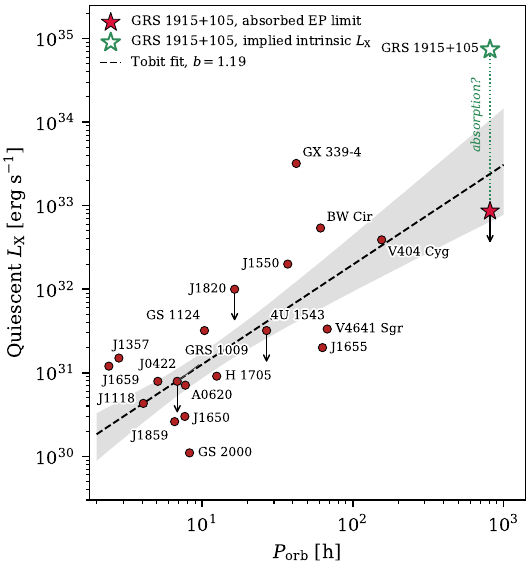}
\caption{Quiescent X-ray luminosity versus orbital period for the sample of Table~\ref{tab:porbsample}. Downward arrows denote $3\sigma$ upper limits. The dashed line and grey band show the censored maximum-likelihood fit, which excludes \grs, and the 16--84\% bootstrap interval on the mean relation. The filled star is \grs, plotted at the absorbed joint \textit{Einstein Probe} limit of 2026 June--August; the open star marks the intrinsic luminosity implied by the radio flux through the radio/X-ray correlation, $\approx$$7\times10^{34}$\,\ergs\ (Sect.~\ref{sec:disc_obscuration}). The thin dotted line labelled ``absorption?'' connecting the two symbols spans the range allowed by the unknown absorption.}
\label{fig:porb_lx}
\end{figure}

\begin{figure}
\centering
\includegraphics[width=\columnwidth]{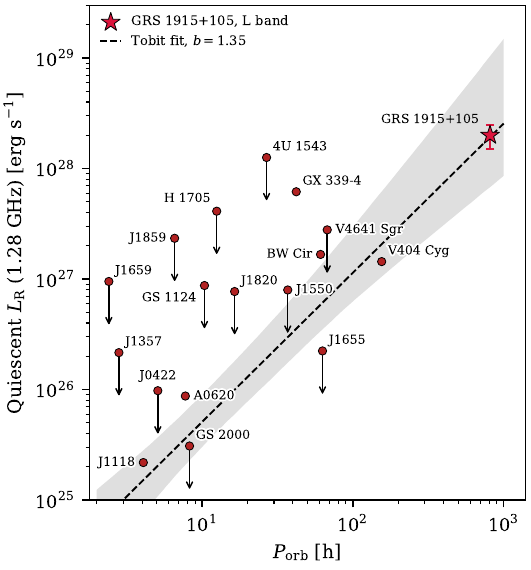}
\caption{Quiescent 1.28\,GHz radio luminosity versus orbital period for the same sample, with $3\sigma$ upper limits recomputed homogeneously from the published flux densities and placed at 1.28\,GHz under the flat-spectrum assumption. The dashed line and grey band show the censored fit excluding \grs\ and the 16--84\% bootstrap interval on the mean relation; the points are expected to scatter about the fit by the intrinsic dispersion of the relation, a factor of $\approx$4 in luminosity, within which nearly all the upper limits are accommodated. The star is \grs: at the native frequency this is the 2026 July L-band measurement directly, with no spectral assumption. }
\label{fig:porb_lr}
\end{figure}

The correlations are shown in Figs.~\ref{fig:porb_lx} and \ref{fig:porb_lr}. For the X-rays we find a censored Kendall $\tau=0.47$ ($p=0.007$) and $L_{\rm X}\propto P_{\rm orb}^{1.19^{+0.34}_{-0.29}}$, with a fitted dispersion about the relation corresponding to a factor of $\approx$4.5 in luminosity. At the orbital period of \grs\ (812.4\,hr; \citealt{Steeghs2013}) the best-fitting relation predicts $L_{\rm X}\simeq2.4\times10^{33}$\,\ergs, with a 16--84\% interval of $3.5\times10^{32}$--$2.0\times10^{34}$\,\ergs from the fit plus the scatter. For the radio we find $\tau=0.63$ ($p=0.04$) and $L_{\rm R}\propto P_{\rm orb}^{1.35^{+0.41}_{-0.27}}$, with a dispersion corresponding to a factor of $\approx$4, predicting $L_{\rm R}\simeq1.9\times10^{28}$\,\ergs, with a 16--84\% interval of $4.2\times10^{27}$--$1.8\times10^{29}$\,\ergs\  from the fit plus the scatter, at the period of \grs. Here and in the following, the central predictions are those of the best-fitting relations, i.e. the dashed lines in Figs.~\ref{fig:porb_lx} and \ref{fig:porb_lr}. The rank statistic and the slope of the radio relation are invariant under the radio frequency convention, which shifts all literature values by the same factor, whereas its normalisation, and hence the comparison with \grs, depends on it.
Both the X-ray upper limit on the luminosity and the radio detection are consistent with the best-fit relations from the sample of quiescent systems we considered. We stress that the X-ray upper limit refers to the observed luminosity, which may be strongly suppressed by local absorption relative to the intrinsic one. The measured radio luminosity, $(2.0\pm0.5)\times10^{28}$\,\ergs\ at 1.28\,GHz ($10^{28.3}$), lies about 4\% above the best-fitting relation, a difference much smaller than the dispersion of the relation (Fig.~\ref{fig:porb_lr}).
We note, however, that the radio fit rests on five detections among seventeen systems (12 are upper limits), hence its normalisation carries the larger uncertainty of the two.

\section{Discussion}
\label{sec:discussion}

\subsection{The nature of the decline}\label{sec:decline_nature}

In this work we report the deepest L- and S-band observations of \grs, obtained in the faintest radio state recorded since its discovery in 1992. 
Such a dimming episode appears to be fundamentally different from other low-flux phases observed in the past, both prior to and during the obscured phase, and is consistent with the final descent to quiescence of the archetypal microquasar. 
The radio emission, the only band among those monitored that is unaffected by the local absorption, has declined to levels never recorded before,
i.e. several times below the faintest radio levels recorded before the current decline ($\sim$0.5--1\,mJy). Since the source has been
undetected in the soft and hard X-rays since 2025 April and September,
respectively (Sect.~\ref{sec:results_lc}), the suppression of the emission is now, for the first time, simultaneous across all the bands in which the source has been observed. Every previous low-flux phase was instead
confined to a single band: the radio-faint episodes of the active decades occurred while the source was X-ray bright, and the X-ray-faint obscured
phase remained radio-bright. Furthermore, the radio decline has been
sustained and essentially monotonic, proceeding from the decay of the 2025 January--February flare through $\sim$1.1\,mJy in 2025 October and $\sim$0.55\,mJy in 2026 January to the present $\sim$0.15\,mJy, with no
re-brightening at any point and no measurable variability across the six
days of the deep July observations (Sect.~\ref{sec:results_ddt}).

The dramatic dimming of \grs\ is not unexpected: hydrodynamic and disc-instability modelling of the system predicted that the outburst that began in 1992 would exhaust the disc mass reservoir after a total duration of $\sim$20--100\,yr, followed by a quiescence of order $10^{4}$\,yr \citep{Truss2006,Deegan2009}. This is fully consistent with the thirty-four-year-long outburst we have observed.
\grs\ is, moreover, the only BH LMXB known to have sustained Roche-lobe-fed, outburst-level (often close to the Eddington level) accretion for over three decades, and the descent into quiescence of so persistent and long-period a system has never been observed. 

However, directly following this dimming is complicated by the presence of Compton-thick absorption, which since 2019 has prevented direct observations of the accretion processes feeding the jets. Hence, the radio emission is currently the only absorption-independent tracer of the accretion activity of the BH. We thus regard the radio decline as the first direct evidence that the central engine itself is fading, rather than merely receding further behind the absorber.

\subsection{The interplay between obscuration and intrinsic fading}\label{sec:disc_obscuration}

Our data allow an independent estimate of how much X-ray flux the obscurer removed while the source was still detectable. Taking the median observed 1--10\,keV luminosity of the MAXI-detected epochs of the obscured phase, $\sim$$4\times10^{36}$\,\ergs, and comparing it with the median of the historical bright-state values compiled by \citet{Bahramian2018}, $\sim$$2.9\times10^{38}$\,\ergs, we obtain a dimming factor of $\sim$70. Assuming that the intrinsic luminosity remained at the historical level, and computing the band-averaged transmission of the 2--12\,keV band for a $\Gamma=2.1$ power law with \citet{Morrison1983} photoelectric cross-sections and Klein--Nishina Compton losses, this corresponds to an effective column density of $N_{\rm H}\approx1.7\times10^{24}$\,cm$^{-2}$, consistent with the Compton-thick threshold $\sigma_{\rm T}^{-1}=1.5\times10^{24}$\,cm$^{-2}$. %
We stress that this is an effective, line-of-sight-averaged column: should a fraction of the observed flux reach us scattered rather than transmitted, the true column would be higher, whereas should the source have already begun to fade intrinsically over this interval (see below) the column would be correspondingly lower.

This estimate, which characterises the earlier and less extreme phase of the obscuration, may be contrasted with the requirements imposed by the deep \textit{NuSTAR} pointing of 2025 November 17, which yielded $L_{3-78}\lesssim1.4\times10^{33}$\,\ergs\ \citep{Gandhi2026}. Those authors show that reproducing such a deficit through absorption by a fully covering absorber, with a scattered fraction below $10^{-4}$, would require a column in excess of $2\times10^{25}$\,cm$^{-2}$, and they further argue that the reprocessed power should in this case emerge in the infrared at levels that the SPHEREx observations would have detected but did not. The obscuring column would hence have had to increase by more than an order of magnitude between the epochs sampled by MAXI and late 2025. We regard a genuine decline in the accretion power as the simpler explanation, which is also what the radio fading independently indicates.

The radio measurements allow us to approach the same question from the opposite direction, i.e. without assuming an intrinsic X-ray luminosity at all. If we suppose that \grs\ lies on the radio/X-ray correlation, then the radio flux predicts the intrinsic X-ray luminosity directly. Adopting the black-hole and black-hole-candidate correlation of \citet{CrookMansour2026} drawn in Fig.~\ref{fig:lrlx} (Sect.~\ref{sec:results_lrlx}), together with the 2026 July L-band measurement $L_{\rm R}=(2.0\pm0.5)\times10^{28}$\,\ergs\ at 1.28\,GHz, we obtain an intrinsic $L_{\rm X}(1$--$10\,{\rm keV})\approx7\times10^{34}$\,\ergs. The uncertainty on this estimate is dominated by the intrinsic scatter of the correlation (a factor of $\approx$3 in $L_{\rm R}$, which corresponds to a factor of $\approx$7 in $L_{\rm X}$ for the slope $\beta=0.57$), so that the implied range is roughly $10^{34}$ to $5\times10^{35}$\,\ergs. %

It follows that our \textit{Einstein Probe} non-detection cannot by itself discriminate between increased absorption and intrinsic fading.
Computing the transmission of the 1--10\,keV band exactly as above, an intrinsic luminosity of $7\times10^{34}$\,\ergs\ would be observed at $\sim$$4\times10^{32}$\,\ergs\ through the $1.7\times10^{24}$\,cm$^{-2}$ derived above, and at a negligible level through any column exceeding $10^{25}$\,cm$^{-2}$. Indeed a column of $\sim$$10^{24}$\,cm$^{-2}$, still below the Compton-thick threshold, would already suffice to hide a source sitting on the correlation at such a low intrinsic luminosity from the joint \textit{Einstein Probe} limit. Hence the soft-band \textit{Einstein Probe} non-detection is consistent with a central engine at the intrinsic luminosity implied by the radio emission ($\sim$$7\times10^{34}$\,\ergs) behind the absorber, and the case for an intrinsic decline rests on the hard band and on the radio rather than on the soft X-rays.

Applied to the hard band, the same reasoning leads to a requirement qualitatively similar to that of \citet{Gandhi2026}, derived from our own data. The nearest core-only radio measurement to their pointing, the jetKAT S-band detection of 2025 December 2 (Sect.~\ref{sec:results_lrlx}), implies an intrinsic $L_{\rm X}(1$--$10\,{\rm keV})\approx3.7\times10^{35}$\,\ergs\ for a source on the correlation, and hence $L(3$--$78\,{\rm keV})\approx4.4\times10^{35}$\,\ergs\ for $\Gamma=2.1$.
Suppressing that below their limit of $1.4\times10^{33}$\,\ergs\ by transmission alone would require $N_{\rm H}\gtrsim7\times10^{24}$\,cm$^{-2}$, a factor of $\sim$4 above the effective column of the MAXI-detected phase, and derived here without assuming that the intrinsic luminosity remained at its historical level. Our value falls below their $>2\times10^{25}$\,cm$^{-2}$ because \citet{Gandhi2026} impose in addition a covering factor of $4\pi$ and a scattered fraction $f_{\rm scatt}<10^{-4}$, whereas we consider transmission alone, so that the two are consistent.

\subsection{Comparison with quiescent black-hole binaries}
\label{sec:disc_quiescence}

Only a handful of BH LMXBs have ever been detected in the radio in the quiescent regime: V404\,Cyg, A0620$-$00, XTE\,J1118$+$480, BW\,Cir, GX\,339$-$4 and MAXI\,J1348$-$630 \citep{Gallo2005,Gallo2006,Gallo2014,Plotkin2021,Tremou2020,Carotenuto2022}. At $L_{\rm R}=(2.0\pm0.5)\times10^{28}$\,\ergs\ (1.28\,GHz; the literature values, measured mostly at 5--8.5\,GHz, are flat-shifted to the same frequency), \grs\ is the most radio-luminous BH LMXB detected in this regime, an order of magnitude above V404\,Cyg (a factor of $\sim$8--14, allowing for the variability of the latter). We note that quiescent compact jets tend to show flat-to-slightly-inverted spectra, so the flat down-shift likely overestimates the 1.28\,GHz luminosities of the comparison systems, which makes this statement conservative.
This is not unexpected, as in BH LMXBs the quiescent luminosity is thought to be set by the residual accretion rate onto the black hole. In quiescence the accretion disc sits on the cold, neutral branch of the disc-instability model \citep{Lasota2001} and most of the transferred mass accumulates in the outer disc, and does not accrete. The residual mass that leaks inward feeds a hot, radiatively inefficient inner flow, which powers the quiescent X-ray and radio emission. How intense the inward leakage is depends on the mass supply of the donor and on the disc size, both of which scale with the size of the binary. 
In \grs\ the donor star is an evolved star, a K--M\,III giant \citep{Greiner2001,Steeghs2013}, transferring mass at a rate of the order of $10^{-8}\,M_{\odot}\,{\rm yr}^{-1}$ \citep{Deegan2009}, far exceeding the typical rates of donor stars in shorter-period binaries (of order $10^{-10}$--$10^{-11}\,M_{\odot}\,{\rm yr}^{-1}$). Hence, the expected quiescent mass accretion rate in \grs\ is most likely significantly higher than in other binaries, even if only a fraction of the mass transferred by the donor is accreted on the BH. This is supported by the case of V404\,Cyg, the most luminous quiescent BH LMXB known prior to \grs, which hosts a K3~III companion of $\sim$0.5--0.7\,M$_{\odot}$ in a 6.47\,d orbit \citep{Casares1992,Khargharia2010}. 
In the disc-instability model the quiescent accretion rate at a given radius must be lower than the local critical value, and this limit rises steeply with radius (roughly $\dot{M} \propto R^{2.6}$, \citealt{Lasota2001}). The inner hot flow is fed at the truncation radius of the cold disc, which is expected to lie at larger radii in larger discs, so that the critical rate there, and hence the maximum quiescent accretion rate onto the BH, is correspondingly higher. With $P_{\rm orb}=33.85$\,d, \grs\ has a binary separation roughly three times larger, and a disc area about an order of magnitude larger, than those of V404\,Cyg, which naturally explains a high quiescent luminosity.

At the orbital period of \grs\ ($P_{\rm orb}=812.4$\,hr), the best fit to the radio luminosity versus the orbital period distribution implies a radio luminosity range of $6.8\times10^{27}$ to $1.0\times10^{29}$\,\ergs\ (16--84\% bootstrap percentiles of the mean relation evaluated at the period of \grs, at 1.28\,GHz). The measured native L-band luminosity, $(2.0\pm0.5)\times10^{28}$\,\ergs, lies about 4\% above the best-fitting relation, $1.9\times10^{28}$\,\ergs. Since the radio luminosity is unaffected by absorption, the 2026 July measurement is fully consistent with the quiescent radio luminosity expected for a BH LMXB of this orbital period.

The X-ray comparison is necessarily one-sided, given that only an absorbed upper limit is available, but it is consistent with the same picture. At the orbital period of \grs\ the best-fitting $L_{\rm X}$--$P_{\rm orb}$ relation predicts $L_{\rm X}\approx2.4\times10^{33}$\,\ergs, with a 16--84\% interval that, once the intrinsic scatter is included, extends up to $\approx$$2.0\times10^{34}$\,\ergs\ (Sect.~\ref{sec:results_porb}). The joint \textit{Einstein Probe} limit, $8.6\times10^{32}$\,\ergs, lies a factor of $\sim$3 below the central prediction but within this interval, hence it is consistent with the quiescent population even in the absence of absorption. If residual absorption suppresses the observed flux by a factor $f$, the intrinsic luminosity corresponding to our limit, $f\times8.6\times10^{32}$\,\ergs, remains within the 84th percentile of the relation for $f\lesssim23$. The $\approx$$7\times10^{34}$\,\ergs\ implied by the radio/X-ray correlation (Sect.~\ref{sec:disc_obscuration}) would instead require $f\sim80$, and it lies a factor of $\sim$3.5 above that envelope, although the two estimates remain consistent once the large intrinsic scatter of the correlation itself is allowed for.

\subsection{The nature of the radio spectrum}
\label{sec:disc_spectrum}

Our deep 2026 July MeerKAT observations yielded a stacked
$\alpha=-0.67\pm0.36\,({\rm stat})\pm0.20\,({\rm sys})$ across
1.28--2.62\,GHz. Taken at face value, this is consistent with optically
thin synchrotron emission, as would be expected from expanding ejecta,
whereas a flat or inverted, self-absorbed spectrum ($\alpha\gtrsim0$), the
canonical signature of a compact jet \citep{Blandford1979}, is mildly
disfavoured at $1.6\sigma$. This is unexpected for a compact jet at such a low luminosity, since quiescent BH LMXBs generally show flat or inverted spectra (see below). Decomposing the two stacked flux densities into
a flat component plus an optically thin one shows that a compact core of
up to $\sim$100\,\ujy\ (95\% confidence, almost independently of
the slope assumed for the thin component) may hide beneath the steep
spectrum, although the best fit prefers a negligible core for thin slopes
of $-0.6$ to $-0.7$. Such a core corresponds to a flat-equivalent
$L_{\rm R}\lesssim1.4\times10^{28}$\,\ergs\ at 1.28\,GHz, which remains
consistent with the lower half of the quiescent range expected for a
binary of \grs's orbital period (Sect.~\ref{sec:results_porb}). Hence, assuming a core jet must show a flat spectrum, our data do not require the compact jet to have vanished. They only require
that it does not dominate at these frequencies, contributing at most
$\sim$70\% of the L-band flux density.

A steep component is nonetheless unexpected in this regime. Every
quiescent BH binary with a well-measured radio spectrum has proven
flat or inverted. V404\,Cyg, the closest analogue to \grs\ in
orbital period, donor class and quiescent luminosity, averages
$\langle\alpha\rangle=0.02$ over 24\,yr of archival measurements
\citep{Plotkin2019}, and its jet spectrum extends with
$\alpha=0.04\pm0.01$ from the radio through the millimetre to $21\,\mu$m
\citep{Borowski2025}; A0620$-$00 is strongly inverted,
$\alpha=+0.74\pm0.19$ \citep{Dincer2018}; and MAXI\,J1820$+$070 remained
at $\alpha=+0.24\pm0.06$ throughout its descent to
$\sim$$4\times10^{-7}L_{\rm Edd}$ \citep{Shaw2021}. The only steep
quiescent index in the literature is that of GX\,339$-$4, $\alpha=-0.8\pm0.4$,
albeit from in-band, sub-band photometry with acknowledged calibration
caveats \citep{Tremou2020}. Additionally, radio-quiet hard-state jets form
a population with mean $\alpha=-0.2$ and width $0.3$, of which our value
sits on the steep tail rather than outside it \citep{Espinasse2018}.
\grs\ itself, in its historical $\sim$5\,mJy radio-faint state, once
returned a formally thin slope, $\alpha=-0.5\pm0.1$ between 5.0 and
15.2\,GHz, although the two frequencies were observed six hours apart and
the flux may have varied in between, whereas the single secure faint-state
measurement, taken a fortnight later, was inverted
\citep[$\alpha=+0.12\pm0.01$;][]{Muno2001}. We consider it unlikely that variability is responsible for our steep $\alpha$, as measurements across epochs are fully consistent, and the across-bands $\alpha$ is consistent with the in-band $\alpha$ from both L- and S-band epochs.

The most economical reading of this spectral slope is that the 2026 July
emission is dominated by optically thin material left over from the
low-level activity that followed the last flaring episode, between 2025
and 2026 (Sect.~\ref{sec:results_lc}). In this regard we note that the
jets launched during the obscured state have been shown to be slow
($\beta\Gamma\lesssim0.4$; \citealt{Yan2026}). Slowly expanding ejecta fade more gradually than fast ones, so that the survival of such material at the $\sim$0.1\,mJy level months after the last activity appears plausible. The steep spectrum, the continuation of the monitoring decline down to the level of the deep observations, and the absence of variability within the campaign are all consistent with this picture. The ejection observed in 2023 April showed a proper motion of approximately 7.8~mas/day (approaching jet) and 6.2~mas/day (receding jet) \citep{Jiang2026}.
Assuming that the ejecta launched in 2025 January \citep{Yan2026} are moving at the same velocity as those observed in 2023, in 2026 July (after $\sim$550\,d) the ejecta may have reached approximately 4.3$''$ and 3.4$''$ (approaching and receding, respectively) in the case of pure ballistic motion. 
Our detections at both L and S band do not show any extension, although it is plausible that unresolved faint ejecta are still present along with a faint core.
In this scenario, the radio luminosity attributable to the compact jet in our
deep images would be at most $L_{\rm R}\lesssim1.4\times10^{28}$\,\ergs,
set by the 100\,\ujy\ upper limit on the flat component, that is, a factor
$\sim$1.5  lower than the value calculated directly from the
total L-band flux density, a shift well within the scatter of the
correlations used in Sects.~\ref{sec:results_porb} and~\ref{sec:disc_obscuration}.

Alternatively, the compact jet may survive with a genuinely steeper
spectrum. A possibility is a transition towards the radio-quiet branch, whose spectral-index distribution extends to values comparable to ours (\citealt{Espinasse2018, Carotenuto2021}), or a jet whose particle acceleration is switching off, for which theory permits steep
radio slopes when re-acceleration along the flow ceases
\citep{Peer2009}. We note, however, that in internal-shock models
the removal of dissipation drives the spectrum towards inversion rather
than towards steepness \citep{Malzac2014}, hence the theoretical
expectation for a dying jet is ambiguous. 

A third possibility, that
the optically-thick-to-thin break of a surviving compact jet has migrated
into or below our band, would require $\nu_{\rm b}\lesssim1.3$\,GHz, about
two orders of magnitude below the lowest break measured in any
steady hard-state jet \citep{Russell2013b,Russell2014,Koljonen2015};
sub-GHz breaks have, however, been observed in quenching and reigniting
jets \citep{Corbel2013,Echiburu2024}, and a jet in the act of switching off is
precisely such a transitional object, so this possibility, though
unlikely, cannot be excluded.

\section{Conclusions}
\label{sec:conclusions}

We have presented the radio and X-ray evolution of \grs\ over the past
few years, covering the final stages of its long-lived, Compton-thick
obscured phase, by combining the ThunderKAT and X-KAT MeerKAT monitoring
at 1.28\,GHz and the AMI-LA record at 15.5\,GHz with deep dual-band
MeerKAT observations and dedicated \textit{Einstein Probe}/FXT pointings.
In 2026 July we measured flux densities of $\sim$150 and $\sim$90\,\ujy\
in L and S band, respectively, the lowest recorded from the system since
its discovery, corresponding to $L_{\rm R}\simeq2\times10^{28}$\,\ergs\
at 1.28\,GHz, together with the first radio spectral index obtained from
this source in this regime, $\alpha=-0.67\pm0.41$.
We argue that \grs\ has likely ended the outburst that began in 1992 and is now
settling into quiescence. This conclusion rests on the following lines of evidence:

\begin{itemize}
\item the radio emission, the only monitored band unaffected by the local absorption, is at present the only accessible tracer of the accretion activity, since the obscuration prevents us from observing the latter directly in the X-rays. This emission has declined to several times below the faintest levels recorded before the current decline, so that the source is now, for the first time, faint in every band in which it has been observed (Sect.~\ref{sec:decline_nature});
\item the radio decline has been steady and essentially monotonic over
 the $\sim$14 months since the end of the last flaring episode (2025 January--May), with no
  re-brightening at any point (Sect.~\ref{sec:results_lc});
\item the source has faded to the luminosities expected of a quiescent black-hole binary of its 33.85\,d orbital period: the measured radio luminosity and the X-ray upper limit are both consistent with the $L_{\rm R}$--$P_{\rm orb}$, $L_{\rm X}$--$P_{\rm orb}$ and $L_{\rm R}$--$L_{\rm X}$ relations traced
  by the quiescent population (Sects.~\ref{sec:results_lrlx}, \ref{sec:results_porb} and
  \ref{sec:disc_quiescence});
\item the disc mass reservoir predicted by hydrodynamic and
  disc-instability modelling should be running out at about this time,
  the predicted outburst duration of $\sim$20--100\,yr being consistent
  with the thirty-four years elapsed \citep{Truss2006,Deegan2009};
\item the alternative, an undiminished central engine hidden behind a
  thicker absorber, appears disfavoured, as it would require the column
  to increase by more than an order of magnitude, to
  $N_{\rm H}\gtrsim2\times10^{25}$\,cm$^{-2}$ for a fully covering absorber, and the reprocessed emission of
  so thick an absorber should have emerged in the infrared, which
  the SPHEREx observations do not show \citep[][Sect.~\ref{sec:disc_obscuration}]{Gandhi2026}.
\end{itemize}

The measured spectral index adds a caveat that operates in one direction
only: should part of the 2026 July flux arise from residual optically
thin emission rather than from a compact core
(Sect.~\ref{sec:disc_spectrum}), the core would be fainter still, which
would strengthen, not weaken, the case for quiescence.

If what we observe is indeed the onset of quiescence in \grs, disc-instability models predict it to last of order $10^{4}$\,yr, so that the transition now under way would not recur on any observationally accessible timescale. Two opportunities follow from this. First, the fading of the accretion-driven emission enables studies of the evolved donor largely free of the disc contamination and the obscuration that have hampered such work since discovery, allowing its spectral type and wind to be constrained more precisely \citep{Greiner2001,Steeghs2013}. 
Second, the switch-off of the core provides a rare test of the origin of the very-high-energy emission associated with \grs\ \citep{LHAASO2025,MartiDevesa2025}: should that emission persist at its current level, its origin at the jet--ISM interaction sites, tens of
parsecs from the black hole, would be strongly favoured, a conclusion potentially extending to microquasars as a class, whereas any fading on a timescale of years, accompanied by a displacement of the fitted centroid towards the extended structures, would instead reveal a contribution from the vicinity of the BH.

\section*{Data availability}

The MeerKAT visibilities are stored in the SARAO archive (\url{https://archive.sarao.ac.za}) under the project codes SCI-20180516-PW-01 (ThunderKAT), SCI-20230907-RF-01 (X-KAT) and DDT-20260703-SM-01 (DDT). The ThunderKAT data and part of the X-KAT data are public, and the remaining data will become public at the end of their proprietary periods. The \textit{Einstein Probe}/FXT data (ObsIDs 06800001589, 06800001667 and 06800001668) will become public through the ESA \textit{Einstein Probe} Science Archive (\url{https://ama.esac.esa.int}) and the Chinese National Astronomical Data Centre after the proprietary period. The MAXI/GSC and \textit{Swift}/BAT light curves are public (\url{http://maxi.riken.jp} and \url{https://swift.gsfc.nasa.gov/results/transients/}), and the AMI-LA 15.5\,GHz data are available from the AMI team on reasonable request. The comparison sample of Appendix~\ref{app:porbsample} is compiled from the published literature.

\begin{acknowledgements}
We thank the anonymous referee, whose comments helped to improve the clarity of the manuscript. SEM thanks SARAO for approving the DDT request DDT-20260703-SM-01, and acknowledges support from the INAF Fundamental Research Grant (2022) EJECTA. RF acknowledges support from the UKRI, European Research Council Synergy Grant 'Blackholistic' (grant agreement number 101071643) and The Hintze Family Charitable Foundation. AM acknowledges support from the Fund Vera Rubin/Chile 2024, under the project DIA 1736 ``Silent black holes around red supergiants'', the H2020 ERC Consolidator Grant ``MAGNESIA'' under grant agreement No. 817661 (PI: Rea) and the National Spanish grant PID2023-153099NA-I00 (PI: Coti Zelati). FCZ is supported by a Ram\'on y Cajal fellowship (grant agreement RYC2021-030888-I) and the National Spanish grant PID2023-153099NA-I00. CS acknowledges funding from the Italian Space Agency, contract ASI/INAF n. I/004/11/6, and financial support from the Bando Ricerca Fondamentale INAF 2024 Large Grant ``Timing the Ultra Luminous X-ray Pulsars (TULiP)''. MCB acknowledges support from the INAF Fundamental Research Grant XBOOM and from the L'Oréal-UNESCO For Women in Science Programme – Italy, through the 2026 Fellowship Award. PA and LDR are supported by the WISE fellowship program, which is financed by NWO.

The MeerKAT telescope is operated by the South African Radio Astronomy Observatory, which is a facility of the National Research Foundation, an agency of the Department of Science and Innovation. This work is based on data obtained with \textit{Einstein Probe}, a space mission supported by the Strategic Priority Program on Space Science of the Chinese Academy of Sciences, in collaboration with ESA, MPE and CNES. We thank the staff of the Mullard Radio Astronomy Observatory for their assistance in the operation of AMI-LA. This research has made use of MAXI data provided by RIKEN, JAXA and the MAXI team, and of \textit{Swift}/BAT transient monitor results provided by the \textit{Swift}/BAT team. We acknowledge the use of the Ilifu cloud computing facility (\url{https://www.ilifu.ac.za}), a partnership between the University of Cape Town, the University of the Western Cape, Stellenbosch University, Sol Plaatje University and the Cape Peninsula University of Technology, supported by the Inter-University Institute for Data Intensive Astronomy (IDIA), the Computational Biology division at UCT and the Data Intensive Research Initiative of South Africa (DIRISA).

\end{acknowledgements}

\bibliographystyle{aa}
\bibliography{biblio}

\appendix
\onecolumn
\section{MeerKAT monitoring log}
\label{app:log}

We list in Table~\ref{tab:log_meerkat} the flux densities of \grs\ measured in the MeerKAT 1.28\,GHz monitoring used in Sects.~\ref{sec:results_lc} and \ref{sec:results_lrlx}.

\begingroup
\small
\setlength{\tabcolsep}{4pt}
\setlength{\LTcapwidth}{\textwidth}
\begin{longtable}{cr@{}lr@{\hspace{3mm}}cr@{}lr@{\hspace{3mm}}cr@{}lr}
\caption{\label{tab:log_meerkat}MeerKAT 1.28\,GHz monitoring of \grs\ obtained within the ThunderKAT and X-KAT programmes between 2018 December 8 and 2026 June 27 (155 epochs). $S_\nu$ is the integrated flux density from the beam-constrained point-source fit, and $\sigma$ is its total uncertainty, which combines in quadrature the larger of the formal fit error and the local image rms with a 5\% flux-scale term (Sect.~\ref{sec:results_lc}). The MJD is that of the \texttt{DATE-OBS} keyword of each image, i.e. the start of the imaged on-source data. The eight epochs marked with $^{\dagger}$ fall below the detection threshold of Sect.~\ref{sec:obs_reduction}, that is, their fitted amplitude is less than three times the local rms, and for these we list the fitted amplitude rather than an upper limit. All epochs, including those presented by \citet{Motta2021}, were re-fitted homogeneously, hence the earlier values may differ slightly from the published ones. The deep DDT observations of 2026 July are listed separately in Table~\ref{tab:imfit}. Epochs are in chronological order, running across each row.}\\
\hline\hline
MJD & \multicolumn{2}{c}{$S_\nu$} & $\sigma$ & MJD & \multicolumn{2}{c}{$S_\nu$} & $\sigma$ & MJD & \multicolumn{2}{c}{$S_\nu$} & $\sigma$ \\
 & \multicolumn{2}{c}{(mJy)} & (mJy) &  & \multicolumn{2}{c}{(mJy)} & (mJy) &  & \multicolumn{2}{c}{(mJy)} & (mJy) \\
\hline
\noalign{\smallskip}
\endfirsthead
\caption{continued.}\\
\hline\hline
MJD & \multicolumn{2}{c}{$S_\nu$} & $\sigma$ & MJD & \multicolumn{2}{c}{$S_\nu$} & $\sigma$ & MJD & \multicolumn{2}{c}{$S_\nu$} & $\sigma$ \\
 & \multicolumn{2}{c}{(mJy)} & (mJy) &  & \multicolumn{2}{c}{(mJy)} & (mJy) &  & \multicolumn{2}{c}{(mJy)} & (mJy) \\
\hline
\noalign{\smallskip}
\endhead
\noalign{\smallskip}
\hline
\endfoot
58460.488 & 2.587 &  & 0.135 & 58686.877 & 66.475 &  & 3.325 & 58691.768 & 42.713 &  & 2.137 \\
58699.848 & 75.823 &  & 3.792 & 58733.715 & 48.460 &  & 2.424 & 58768.771 & 15.546 &  & 0.781 \\
58910.208 & 8.612 &  & 0.438 & 58942.195 & 122.223 &  & 6.112 & 58957.144 & 19.584 &  & 0.982 \\
58964.201 & 120.240 &  & 6.012 & 58971.020 & 88.600 &  & 4.431 & 58978.174 & 42.460 &  & 2.124 \\
58986.174 & 119.295 &  & 5.965 & 58992.021 & 2.223 &  & 0.420 & 59001.110 & 17.071 &  & 0.856 \\
59008.156 & 23.737 &  & 1.190 & 59014.907 & 34.937 &  & 1.749 & 59019.073 & 34.336 &  & 1.719 \\
59027.022 & 35.493 &  & 1.776 & 59034.846 & 32.865 &  & 1.645 & 59042.836 & 73.529 &  & 3.677 \\
59049.840 & 643.564 &  & 32.179 & 59054.914 & 236.436 &  & 11.822 & 59062.730 & 255.891 &  & 12.795 \\
59069.719 & 587.048 &  & 29.353 & 59076.793 & 80.969 &  & 4.049 & 59083.906 & 45.611 &  & 2.281 \\
59091.708 & 12.230 &  & 0.616 & 59096.886 & 11.599 &  & 0.585 & 59104.727 & 11.445 &  & 0.576 \\
59112.762 & 7.506 &  & 0.379 & 59118.799 & 14.230 &  & 0.716 & 59125.609 & 8.794 &  & 0.449 \\
59132.673 & 16.032 &  & 0.803 & 59139.542 & 14.897 &  & 0.747 & 59147.584 & 38.031 &  & 1.903 \\
59152.541 & 164.861 &  & 8.244 & 59161.454 & 286.207 &  & 14.315 & 59168.664 & 86.018 &  & 4.302 \\
59175.597 & 59.780 &  & 2.990 & 59181.612 & 34.135 &  & 1.708 & 59188.507 & 66.595 &  & 3.332 \\
59196.383 & 67.982 &  & 3.402 & 59203.529 & 52.926 &  & 2.647 & 59211.421 & 26.771 &  & 1.340 \\
59218.613 & 230.312 &  & 11.518 & 59223.380 & 126.503 &  & 6.326 & 59230.345 & 66.255 &  & 3.314 \\
59238.498 & 44.734 &  & 2.239 & 59244.218 & 45.389 &  & 2.272 & 59251.457 & 215.073 &  & 10.772 \\
59260.443 & 77.289 &  & 3.866 & 59265.467 & 34.087 &  & 1.706 & 59273.326 & 16.428 &  & 0.824 \\
59279.235 & 16.899 &  & 0.848 & 59286.124 & 19.384 &  & 0.971 & 59293.165 & 16.944 &  & 0.849 \\
59301.305 & 16.659 &  & 0.836 & 59309.214 & 27.713 &  & 1.387 & 59314.122 & 32.742 &  & 1.639 \\
59685.223 & 2.224 &  & 0.130 & 60042.237 & 404.809 &  & 20.241 & 60070.186 & 122.109 &  & 6.106 \\
60078.991 & 157.871 &  & 7.895 & 60100.997 & 85.065 &  & 4.254 & 60169.744 & 1363.344 &  & 68.168 \\
60174.674 & 768.501 &  & 38.427 & 60191.673 & 252.970 &  & 12.649 & 60195.652 & 246.844 &  & 12.343 \\
60203.655 & 183.920 &  & 9.197 & 60218.561 & 418.899 &  & 20.947 & 60223.634 & 384.948 &  & 19.248 \\
60239.507 & 399.819 &  & 20.000 & 60245.543 & 375.338 &  & 18.769 & 60254.458 & 1102.537 &  & 55.129 \\
60260.503 & 254.406 &  & 12.721 & 60266.493 & 165.927 &  & 8.297 & 60273.482 & 158.505 &  & 7.926 \\
60280.472 & 711.288 &  & 35.565 & 60288.362 & 172.815 &  & 8.643 & 60295.386 & 196.821 &  & 9.842 \\
60301.348 & 834.172 &  & 41.710 & 60316.353 & 245.965 &  & 12.300 & 60331.257 & 386.127 &  & 19.308 \\
60344.253 & 78.136 &  & 3.908 & 60359.210 & 85.463 &  & 4.274 & 60373.179 & 61.327 &  & 3.068 \\
60385.151 & 52.256 &  & 2.615 & 60393.106 & 54.229 &  & 2.714 & 60400.071 & 45.948 &  & 2.299 \\
60413.071 & 37.581 &  & 1.881 & 60429.068 & 28.082 &  & 1.406 & 60441.988 & 25.344 &  & 1.270 \\
60448.998 & 22.989 &  & 1.151 & 60455.991 & 19.446 &  & 0.976 & 60469.939 & 17.140 &  & 0.859 \\
60471.862 & 4.661 &  & 0.245 & 60476.907 & 15.799 &  & 0.796 & 60490.822 & 13.286 &  & 0.669 \\
60499.831 & 14.849 &  & 0.746 & 60506.795 & 16.048 &  & 0.807 & 60518.811 & 17.323 &  & 0.870 \\
60527.757 & 16.055 &  & 0.808 & 60539.730 & 10.841 &  & 0.547 & 60546.645 & 8.233 &  & 0.426 \\
60553.710 & 1.609 &  & 0.126 & 60554.627 & 6.702 &  & 0.355 & 60569.643 & 4.903 &  & 0.253 \\
60574.579 & 4.237 &  & 0.224 & 60597.553 & 3.686 &  & 0.208 & 60603.574 & 3.720 &  & 0.194 \\
60624.437 & 3.961 &  & 0.216 & 60636.482 & 14.917 &  & 0.749 & 60652.399 & 8.919 &  & 0.454 \\
60658.343 & 5.890 &  & 0.307 & 60665.324 & 4.112 &  & 0.228 & 60673.339 & 8.678 &  & 0.438 \\
60700.233 & 1075.799 &  & 53.791 & 60716.193 & 759.399 &  & 37.973 & 60728.164 & 1224.299 &  & 61.215 \\
60737.206 & 603.707 &  & 30.186 & 60751.159 & 225.308 &  & 11.267 & 60764.179 & 253.913 &  & 12.696 \\
60777.155 & 213.621 &  & 10.682 & 60793.112 & 80.723 &  & 4.038 & 60806.035 & 32.296 &  & 1.618 \\
60813.933 & 21.985 &  & 1.107 & 60817.948 & 22.188 &  & 1.112 & 60835.963 & 10.811 &  & 0.545 \\
60848.842 & 7.262 &  & 0.370 & 60862.855 & 5.570 &  & 0.292 & 60874.772 & 4.111 &  & 0.226 \\
60890.771 & 3.215 &  & 0.187 & 60905.678 & 2.250 &  & 0.142 & 60918.642 & 2.009 &  & 0.130 \\
60934.636 & 1.754 &  & 0.103 & 60953.541 & 1.134 &  & 0.128 & 60967.532 & 1.106 &  & 0.119 \\
60995.435 & 0.918 &  & 0.121 & 61021.408 & 0.679 &  & 0.118 & 61037.380 & 0.765 &  & 0.154 \\
61052.287 & 0.554 &  & 0.091 & 61067.468 & 0.397 &  & 0.071 & 61080.241 & 0.469 &  & 0.078 \\
61093.176 & 0.313 & $^{\dagger}$ & 0.150 & 61106.139 & 0.234 & $^{\dagger}$ & 0.120 & 61120.168 & 0.230 & $^{\dagger}$ & 0.094 \\
61142.084 & 0.424 &  & 0.126 & 61155.011 & 0.329 &  & 0.093 & 61169.015 & 0.361 &  & 0.119 \\
61184.917 & 0.222 & $^{\dagger}$ & 0.119 & 61198.863 & 0.229 & $^{\dagger}$ & 0.155 & 61202.866 & 0.305 & $^{\dagger}$ & 0.104 \\
61217.858 & 0.178 & $^{\dagger}$ & 0.092 & 61218.932 & 0.073 & $^{\dagger}$ & 0.051 & & & & \\
\end{longtable}
\endgroup

\clearpage

\section{The quiescent comparison sample}
\label{app:porbsample}

\begin{table*}[!ht]  %
\centering
\caption{Quiescent luminosities, orbital periods and adopted distances of the transient BH LMXBs used in Sect.~\ref{sec:results_porb}. Radio luminosities are quoted at 1.28\,GHz, the convention of this paper: the literature values, recomputed homogeneously from the published flux densities (measured mostly at 5--8.5\,GHz; 1.28\,GHz for MAXI\,J1348$-$630), are placed at this frequency under the flat-spectrum assumption, whereas the \grs\ value is the native 2026 July L-band measurement of Sect.~\ref{sec:results_lrlx} and involves no spectral assumption. X-ray luminosities are as published (bands range from 0.5--10 to 3--9\,keV, an inhomogeneity of at most a factor of $\sim$2 for $\Gamma\approx2$). All luminosities are quoted at the distance $d$ listed, which is the one adopted by the corresponding reference except for V4641\,Sgr (see below). The \grs\ X-ray value is an absorbed upper limit.}
\label{tab:porbsample}
\begin{tabular}{lccccl}
\hline\hline
System & $P_{\rm orb}$ & $d$ & $L_{\rm R}$ (1.28\,GHz) & $L_{\rm X}$ & Ref. \\
 & (hr) & (kpc) & (erg\,s$^{-1}$) & (erg\,s$^{-1}$) & \\
\hline
\grs                 & 812.4 & 9.4  & $(2.0\pm0.5)\times10^{28}$ & $<8.6\times10^{32}$ & This work; S13; R23 \\
V404\,Cyg            & 155.3 & 2.39 & $1.4\times10^{27}$ & $3.9\times10^{32}$  & C08 \\
V4641\,Sgr           & 67.6  & 6.2  & $<2.8\times10^{27}$ & $3.3\times10^{31}$  & T03; MJ11; MacD14 \\
GRO\,J1655$-$40      & 62.9  & 3.2  & $<2.2\times10^{26}$ & $2.0\times10^{31}$  & G01; MJ11; C10 \\
BW\,Cir              & 61.1  & 7.1  & $1.7\times10^{27}$ & $5.4\times10^{32}$  & P21 \\
GX\,339$-$4          & 42.1  & 8.0  & $6.2\times10^{27}$ & $3.2\times10^{33}$  & T20; He17 \\
XTE\,J1550$-$564     & 37.0  & 4.4  & $<8.0\times10^{26}$ & $2.0\times10^{32}$  & C06; C10 \\
4U\,1543$-$47        & 26.8  & 9.0  & $<1.3\times10^{28}$ & $<3.2\times10^{31}$ & G01; H03; XK \\
MAXI\,J1820$+$070      & 16.4  & 2.96 & $<7.7\times10^{26}$ & $<1.0\times10^{32}$ & XK; To20 \\
H\,1705$-$250        & 12.5  & 8.6  & $<4.1\times10^{27}$ & $9.1\times10^{30}$  & MJ11; Y12 \\
Swift\,J1727.8$-$1613$^{(a)}$ & 10.80 & 3.7 & -- & -- & MS25 \\
GS\,1124$-$684       & 10.4  & 4.95 & $<8.7\times10^{26}$ & $3.2\times10^{31}$  & P21; S02; W16 \\
MAXI\,J1305$-$704$^{(a)}$ & 9.46 & 7.5 & -- & -- & MS21 \\
GS\,2000$+$25          & 8.26  & 2.7  & $<3.1\times10^{25}$ & $1.1\times10^{30}$  & R20 \\
A0620$-$00          & 7.75  & 1.06 & $8.7\times10^{25}$ & $7.1\times10^{30}$  & G06; D18 \\
XTE\,J1650$-$500     & 7.69  & 2.6  & --                  & $3.0\times10^{30}$  & G08 \\
MAXI\,J1803$-$298$^{(a)}$ & 7.0 & 8.0 & -- & -- & MS22 \\
GRS\,1009$-$45       & 6.85  & 5.0  & --                  & $<7.9\times10^{30}$ & H03 \\
XTE\,J1859$+$226       & 6.58  & 6.3  & $<2.3\times10^{27}$ & $2.6\times10^{30}$  & T03; MJ11; CS11 \\
GRO\,J0422$+$32        & 5.09  & 2.6  & $<9.8\times10^{25}$ & $7.9\times10^{30}$  & G01; MJ11 \\
XTE\,J1118$+$480       & 4.08  & 1.72 & $2.2\times10^{25}$ & $4.3\times10^{30}$  & G14 \\
Swift\,J1753.5$-$0127$^{(a)}$ & 3.24 & 3.9 & -- & -- & Z08; P17 \\
Swift\,J1357.2$-$0933 & 2.81 & 6.3  & $<2.2\times10^{26}$ & $1.5\times10^{31}$  & P16 \\
MAXI\,J1659$-$152    & 2.41  & 6.0  & $<9.5\times10^{26}$ & $1.2\times10^{31}$  & H13; J12 \\
MAXI\,J1348$-$630$^{(b)}$ & --  & 2.2  & $1.1\times10^{26}$ & $7.5\times10^{30}$  & C22 \\
\hline
\end{tabular}
\tablefoot{Upper limits are $3\sigma$. $^{(a)}$Systems satisfying the period criterion but without published quiescent luminosities, listed for completeness and excluded from the fits; Swift\,J1753.5$-$0127's lowest published radio/X-ray pairs \citep{Plotkin2017} were obtained on the rise out of quiescence and are not quiescent levels. $^{(b)}$No measured orbital period (only a $\lesssim$10--20\,hr estimate exists); the system appears on the $L_{\rm R}$--$L_{\rm X}$ plane but not in the $P_{\rm orb}$ fits. For V4641\,Sgr we adopt the refined distance of $6.2\pm0.7$\,kpc of \citet{MacDonald2014} in place of the 9.6\,kpc assumed by the original references, which lowers both its luminosities by a factor of $\sim$2.4. Further distance caveats: the BW\,Cir Gaia EDR3 distance conflicts with the $>$25\,kpc donor-based estimate \citep{Plotkin2021}, and Swift\,J1357.2$-$0933 may lie at 2.3\,kpc rather than 6.3\,kpc \citep{Plotkin2016}, lowering both its luminosities by a factor $\sim$7.5.
References: S13 \citep{Steeghs2013}; R23 \citep{Reid2023}; C08 \citep{Corbel2008}; T03 \citep{Tomsick2003}; MJ11 \citep{Miller-Jones2011}; G01 \citep{Garcia2001}; C10 \citep{Calvelo2010}; P21 \citep{Plotkin2021}; T20 \citep{Tremou2020}; He17 \citep{Heida2017}; C06 \citep{Corbel2006}; H03 \citep{Hameury2003}; XK: unpublished X-KAT MeerKAT limits; To20 \citep{Torres2020}; Y12 \citep{Yang2012}; MS21, MS22, MS25 \citep{MataSanchez2021,MataSanchez2022,MataSanchez2025}; MacD14 \citep{MacDonald2014}; S02 \citep{Sutaria2002}; W16 \citep{Wu2016}; R20 \citep{Rodriguez2020}; G06 \citep{Gallo2006}; D18 \citep{Dincer2018}; G08 \citep{Gallo2008}; CS11 \citep{Corral-Santana2011}; G14 \citep{Gallo2014}; Z08 \citep{Zurita2008}; P17 \citep{Plotkin2017}; P16 \citep{Plotkin2016}; H13 \citep{Homan2013}; J12 \citep{Jonker2012}; C22 \citep{Carotenuto2022}.}
\end{table*}

\end{document}